\documentclass[traditabstract]{aa} 
\usepackage{graphicx}
\usepackage{natbib}
\usepackage{txfonts}
\usepackage{mathrsfs}

\newcommand{\hd}{\hbox{HD~174884 }}
\newcommand{\hk}{\hbox{HD~174884}}
\newcommand{\corot}{CoRoT}

\newcommand{\ANG}{\accent'27A}

\newcommand{\kms}{\mathrm{km\,s}^{-1} }
\newcommand{\ks}{km\,s$^{-1}$}

\newcommand{\ms}{M$_{\odot}$}
\newcommand{\rs}{R$_{\odot}$}

\bibpunct{(}{)}{;}{a}{}{,} 
\begin{document}

\title{HD 174884: a strongly eccentric, short-period early-type binary system discovered by CoRoT\thanks{Based on photometry collected by the CoRoT space mission and spectroscopy obtained with the CORALIE spectrograph attached to the 1.2m Euler telescope at ESO, La Silla, Chile. The CoRoT space mission was developed and is operated by the French space agency CNES, with participation of ESA's RSSD and Science Programs, Austria, Belgium, Brazil, Germany and Spain}}

   \author{
   C. Maceroni\inst{1}, J. Montalb\'{a}n\inst{2}, E. Michel\inst{3}, P. Harmanec\inst{4},
   A. Pr\u{s}a\inst{5}\fnmsep\inst{6}, M. Briquet\inst{7}\thanks{Postdoctoral fellow of the Fund for Scientific
Research of Flanders (FWO)}, E. Niemczura\inst{7}\fnmsep \inst{8}, T. Morel\inst{2}\fnmsep \inst{7},  D.~Ladjal\inst{7}, 
M. Auvergne\inst{3}, A. Baglin\inst{3}, F. Baudin\inst{9}, C. Catala\inst{3}, R. Samadi\inst{3}, 
          \and
          C. Aerts\inst{7}\fnmsep \inst{10}
          }		  
   \institute{
   INAF - Osservatorio Astronomico di Roma, via Frascati 33, I-00040 Monteporzio C. (RM), Italy\\ 
              \email{maceroni@oa-roma.inaf.it}
   \and Institut d'Astrophysique et G\'{e}ophysique Universit\'{e} de Li\`{e}ge,
			All\'{e}e du 6 A\^{out}, B-4000 Li\`{e}ge, Belgium 
   \and Observatoire de Paris, LESIA, UMR 8109, F-92195 Meudon, France
   \and Astronomical Institute of the Charles University, Faculty of Mathematics and Physics,
      V~Hole\v{s}ovi\v{c}k\'ach 2, CZ-180~00 Praha 8, Czech Republic
   \and Villanova University, Dept.Astron.Astrophys., 800 E Lancaster Ave., Villanova, PA 19085, USA
   \and University of Ljubljana, Dept. of Physics, Jadranska 19, SI-1000, Ljubljana, Slovenia
   \and Instituut for Sterrenkunde, K.U.Leuven, Celestijnenlaan 200 D, B-3001, Leuven, Belgium
   \and Astronomical Institute of Wroc\l aw University, ul. Kopernika 11, 51-622 Wroc\l aw, Poland
   \and Institut dÕAstrophysique Spatiale, Campus dÕOrsay, F-91405 Orsay, France
   \and Dept. of Astrophysics, IMAPP, Radboud University Nijmegen, PO Box 9010, 6500 GL Nijmegen,
   the Netherlands.
   }

   \date{Received September 16, 2009; accepted October 17, 2009}

\abstract{

Accurate photometric CoRoT space observations of a secondary seismological target, \hk, led to the discovery that this star is an astrophysically important double-lined eclipsing spectroscopic
binary in an eccentric orbit ($e\sim0.3$), unusual for its short
3\fd65705 orbital period. The high eccentricity, coupled with the orientation of the binary orbit in space, explains the very unusual observed light curve with strongly unequal primary and secondary eclipses having the depth ratio of 1-to-100 in the CoRoT ``seismo" passband. Without the high accuracy of the CoRoT photometry, the secondary eclipse, 1.5 mmag deep, would  have gone unnoticed. A spectroscopic follow-up program   provided  45 high dispersion spectra. The  analysis of the CoRoT light curve was performed with an adapted version of PHOEBE that supports CoRoT passbands. The final solution was obtained by simultaneous fitting of the light and the radial velocity curves. Individual star spectra were derived by  spectrum disentangling. The uncertainties of the fit were derived by bootstrap resampling and the solution uniqueness was tested by heuristic scanning.

The results provide a consistent picture of the system composed of two late B stars.  The Fourier analysis of the light curve fit residuals yields two components, with orbital frequency multiples and an amplitude of $\sim$ 0.1 mmag, which are tentatively interpreted as tidally induced pulsations.

An extensive  comparison with theoretical models is carried out by means of the Levenberg-Marquardt minimization technique and the discrepancy between models  and the derived parameters is discussed.  The best fitting models yield a young system age of 125 million years which is consistent with the eccentric orbit and  synchronous component rotation at periastron.\\}


    \authorrunning{C. Maceroni et al.} 

    \maketitle 
%

\section{Introduction}
\object{HD 174884} (Corot 7758) was selected as a  seismology target for   the first  short run in the CoRoT 
``center"  direction (SRc1) because  -- on the basis of its B8~V spectral classification  -- it was considered  a 
good candidate to pinpoint the instability strip red edge of the Slowly Pulsating B stars \citep[e.g.][]{wae98}.
Its only available spectrum, obtained before 
launch with the Elodie spectrograph and available from the Gaudi database\footnote{GAUDI (sdc.laeff.inta.es/gaudi/) 
is the data archive  of the ground-based asteroseismology programme of the CoRoT mission. The GAUDI system is 
maintained at LAEFF, which is part of the Space Science Division of INTA.}  \citep{sol05} did not reveal the star's binary
nature, which was instead immediately evident from  CoRoT photometry. 
The system was continuously observed  during 27 days. The light curve of HD~174884 is characterized by signatures of high eccentricity: the secondary minimum markedly displaced from phase 0.5 and a  bump, typical of close 
eccentric binaries, due to variable distortion of stellar surfaces along the orbit. The system is very interesting 
for several reasons: the high eccentricity is coupled with quite short a period ($P_{orb}=3 \fd 66$), 
putting the system in an extreme location of the period - eccentricity diagram of B stars. This indicates
young age or an inefficient circularization process. Secondly, the huge difference between the primary and  secondary minima, the latter detected only thanks to CoRoT accuracy, and the overall shape and accuracy of the light curve is a challenge for light curve fitting methods and a test of their adequacy to solve light curves of this unprecedented quality. Finally, only a thorough analysis of HD~174884 could yield information of the possible pulsational properties of the components.
   \begin{figure*}[!ht]
   \centering

  \includegraphics[width=17.4cm]{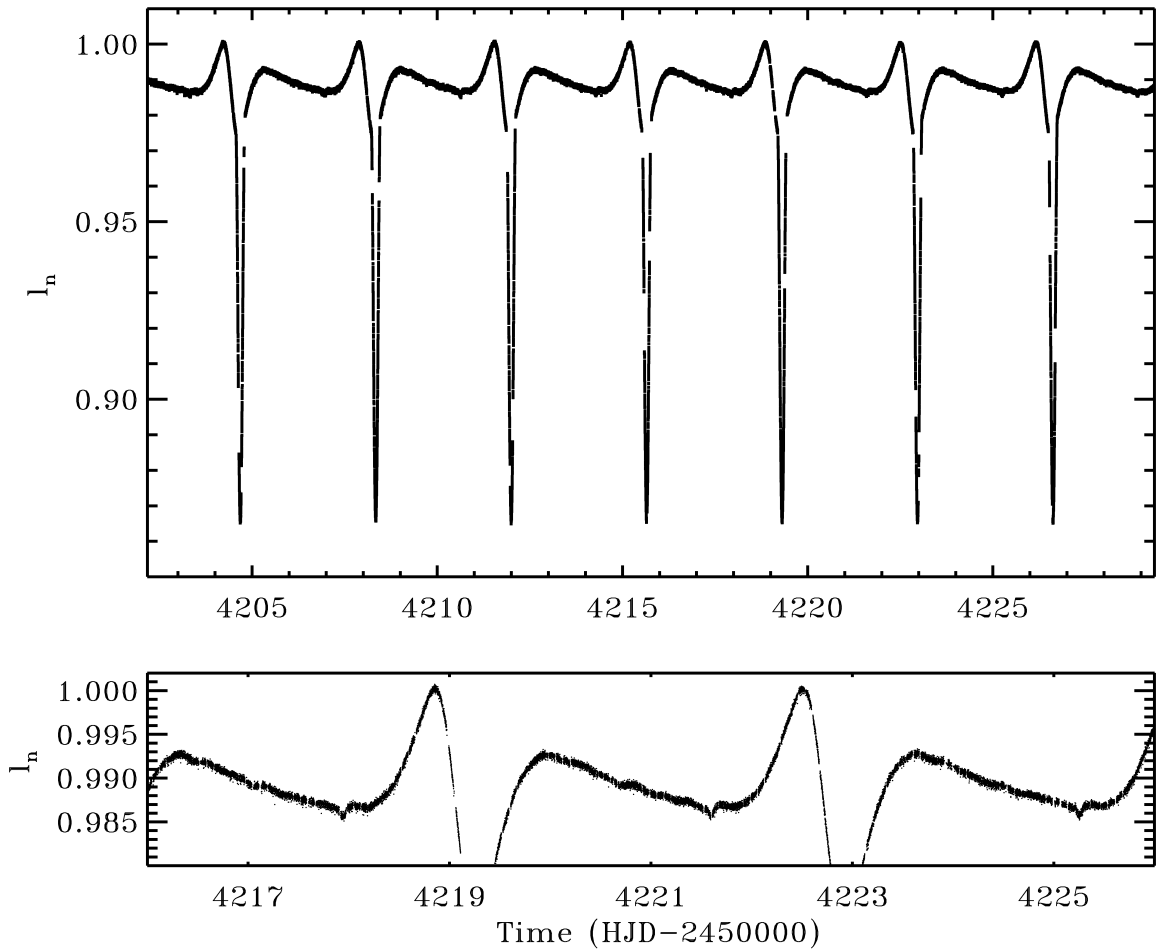}  
      \caption{Upper panel: the complete CoRoT light curve of HD 174884 as normalized flux vs HJD. Lower panel: a 
blow-up, spanning approximately three periods, showing the tiny secondary minimum (first occurrence at reduced HJD of around 4218)  
              }
      \label{lc}
   \end{figure*}
In the following sections we present the results of the analysis of CoRoT photometry and of ground-based high 
dispersion spectroscopy. A critical analysis of parameter uncertainties and of the uniqueness of the light and 
radial velocity curve  solution is an essential part of the study.  

The well constrained physical parameters of the binary components are then compared with  evolutionary models 
computed with  the CL\'ES code \citep[Code Li\`{e}geois d' \'{E}volution Stellaire,][]{scuetal08} with the aim of  
constraining the age and the structural parameters (chemical composition, overshooting). 
Finally the light curve fit residuals, with variations of a few $10^{-4}$ mag, are analyzed.


\section{Observations of HD 174884 and their analysis}
\label{obs}
\subsection{CoRoT photometry and extant data}
\label{phot}
 HD 174884 is a relatively bright star and  several photometric  measurements  could be found in databases and 
catalogs. 
 A spectral classification as B8~V appears in the Simbad database (operated at CDS, Strasbourg, France) and was 
 confirmed by  the Gaudi classification. 
  Available magnitudes and colors are collected in Table \ref{mags}, the Str\"{o}mgren  photometry is from the Gaudi 
database, the infrared ones from 2MASS \citep{2mass}. These magnitudes and colors were dereddened  according to 
\citet{Moon85}, assuming a ratio of total to selective absorption $A_{\mathrm{v}}/E(b-y)=4.28$ 
\citep{crawman76}. For the 2MASS colors the respective ratios were taken from \citet{riele85}.
The  colors, once corrected for the rather heavy absorption, correspond to the expected values for the (primary) 
spectral type, and in particular rule out the presence of a near IR excess,which was suggested by the  observed 
values. 
The star was also observed by Hipparcos, but its parallax is of little use, being affected by a large error,
$\pi =1.95 \pm 0.73 $  mas, according to the  new reduction of the Hipparcos data \citep{vanle07}.

\begin{table}[!hb]
\caption{Colors and magnitudes of HD 174884}             
\label{mags}      
\centering                          
\begin{tabular}{c c c }        
\hline\hline                 
 & Observed & Dereddened \\    
\hline                        
V    	& 7.998$\pm$0.002 & 7.001 $\pm$0.006\\      
b-y  	& 0.173$\pm$0.001 & -0.055$\pm$0.002\\
m$_{1}$	& 0.042 $\pm 0.001$ & 0.117$\pm$0.001\\
c$_{1}$ & 0.637$\pm$0.005 & 0.594$\pm$0.005\\
$\beta$	& 2.701 $\pm$0.011  &       \\ 
J    	& 7.577$\pm $ 0.024  & 7.302$\pm$0.025\\ 
H    	& 7.562$\pm $ 0.023  & 7.391 $\pm $ 0.024\\      
K    	& 7.521$\pm $ 0.021  & 7.412$\pm$0.021\\      
\hline                                   
\end{tabular}
\end{table}

CoRoT observed HD~174884 (Corot ID: 7758) for 27 days\footnote{ The data can be retrieved from the CoRoT public archive  at  http://idoc-corot.ias.u-psud.fr/}. The almost uninterrupted time series  extends  over about 
seven orbital cycles with a  time sampling of $32^{\mathrm{s}}$.
The full light curve has 60016 points, after removal of the measurements  perturbed by South Atlantic Anomaly  crossing (proton events).
Each point is the mean made of 32 1-s exposures and has, therefore, an associated r.m.s. deviation, which is typically of the order of $10^{-3}$ times the measured flux. 

A slow decreasing trend, presumably due to some kind of instrumental decay or gradual pointing drift, was corrected by fitting and subtracting a second order polynomial passing through the maxima. Furthermore, we deleted a few hundred spurious points that arise from the inadequate linear fit used in the first version of the CoRoT pipeline. The complete, normalized, light curve (59492 points) is  shown in Fig.~\ref{lc} together with a blow-up of the central part, displaying a tiny ($\simeq 1.5$ mmag deep) secondary minimum.

\begin{figure}
   \centering
   \includegraphics[width=8.7cm]{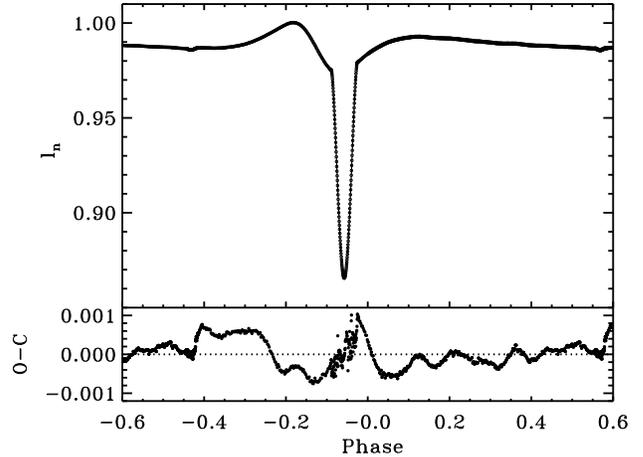}
	
      \caption{Upper panel: the phased and binned CoRoT light curve of HD 174884. Dots: normal points as described in the text. 
The solid line is the light curve fit according to the model of Sect.~\ref{fin_anal}, which can be distinguished from  normal 
points  only during  primary eclipse. Lower panel: the corresponding residuals (observed - computed normalized flux)   
              } 
    \label{phased_lc}
   \end{figure}

The  binary period  was derived in two steps: an initial estimate with   Period04 \citep{p04} 
was refined with the Phase Dispersion Minimization method \citep{stell78}, which is better suited for the analysis of non-sinusoidal
variations. The epoch of primary minimum was derived by a linear fit to the seven minimum epochs, which were obtained 
by parabolic fit to the curve minima. The results were  checked with PHOEBE \citep{prsazw05}. 
 Note that PHOEBE determines
the epoch of zero phase with respect to the argument of periastron (rather than with the position of the primary minimum), and provides phases for important positions in orbit, like superior conjunction\footnote{In the case of an elliptical orbit, the minimum light epoch can be slightly shifted from the superior conjunction; in our case we observe a shift of $\sim 4$ minutes.}. We adopt the following ephemeris:
\begin{equation}
  T_{\mathrm{minI}} = 2454204.68162 (4) +   3 \fd 65705 (1) \times E.
\label{ephe}
\end{equation}
The light curve was  folded with this period and found to be stable within the observed time frame, as the dispersion of the phased curve was comparable to that of the original time series. Therefore, we binned the oversampled curve into 650 normal points, to facilitate the light curve solution. The binning was performed with a variable step, 
$\Delta \phi = 0.002$ outside the eclipses and $\Delta \phi = 0.0005$ during the eclipses to keep good sampling. The mean number of points contributing to normals is $\sim 130$ outside the eclipse and $\sim 30$ inside, the folded light curve is shown in Fig.~\ref{phased_lc}.
\subsubsection{Preliminary light curve analysis} 
\label{prel_lc_anal}

The light curve fit procedure went through several stages, also because the spectroscopic data were obtained only 
some months after the CoRoT run, thanks to  a follow-up program organized after the discovery of binarity. Our final fit 
is a simultaneous solution of both light and radial velocity curves, but the first  steps, based on photometry 
alone, were quite instructive. They were  a test of what can be derived from a single light curve which, on the one side, 
is of unprecedented accuracy but on the other has features suggesting weak constraining of the solution. 

All light curve solutions were performed with PHOEBE \citep{prsazw05}, which was expanded to include  flux computation in the CoRoT passbands for both seismology and exo-planet field. Specific passband intensities and limb darkening coefficients were computed for CoRoT passbands from \citet{castelli2004}'s NEWODF models spanning from 3500\,K to 50,000\,K across the entire H-R diagram. The spectral energy distribution functions (SEDs) were synthesized with SPECTRUM\footnote{http://www.phys.appstate.edu/spectrum/spectrum.html} 2.75 \citep{grayco94} at a 1~\AA\ dispersion, on a 3000~\AA\ to 10,000~\AA\  wavelength range. The limb darkening tables were computed by synthesizing SEDs in 20 values of $\mu = \cos \theta$ from the center of the disk to the limb, and fitting the linear cosine law and non-linear log and square root laws by least squares.

For the preliminary  solutions, we profited from the scripter version of PHOEBE and used Nelder \& Mead's Simplex
minimization for quick descent towards the minima and standard differential corrections when close to the solution.
 The value of primary effective temperature was fixed to that provided  by  Gaudi, which is derived by means of  the
 code TempLogG\footnote{http://www.univie.ac.at/asap/templogg/main.php} and Str\"{o}mgren photometry. TempLogG uses the grids of \citet{moondwo85} in the $T_{\rm eff}$-log g parameter space, with the improvements by \citet{napal93}. The resulting value  is $ T_{\mathrm{eff,1}}=13140$~K; we did not use the internal error from TempLogG but assumed an uncertainty  of 1500~K,  expressing the accuracy of $T_{\rm eff}$ determination for B-stars by different methods \citep{deridetal04}.  For the secondary star we actually started from a $T{\rm eff}$  value (3500~K) very far from the final one (wrongly) assuming, on the basis of the great difference of in-eclipse depths,  a small, cool secondary star. 

We computed, as starting value of a grid,  pairs of eccentricity ($e$) and longitude of periastron ($\omega$) providing the observed distance between maxima. The parameters  adjusted in the minimizations  were inclination $i$, $e$, $\omega$, $T_{\mathrm{eff,2}}$, and surface potentials $\Omega_{1,2}$; the primary passband luminosity $L_1$ was separately computed, rather than adjusted,  as this allows a smoother convergence to the minimum \citep{prsazw05}. For limb darkening we adopted a square root law that employs two coefficients, $x_{LD}$ and $y_{LD}$ per star and per passband. The gravity darkening and albedo coefficients were kept fixed at their theoretical values, $g = 1.0$ and $A = 1.0$ for radiative envelopes \citep{vonzeipel1924}. The mass ratio was set to values between 0.1 and 0.5. When the adjustment of these parameters did not provide further improvement, we attempted to adjust synchronicity parameters (ratios of rotational to orbital velocity), $F_{1,2}$.
  
In spite of the very shallow secondary eclipse and the low inclination, the solution we obtained is fairly close to 
the final one, implying that the single light curve is already  constraining the solution (this was confirmed later on 
by the results of heuristic scan, presented in Section \ref{heuristic_scan}). In particular, we could exclude solutions 
with a cool secondary, in favour of a system configuration  yielding  a grazing eclipse of a relatively hot secondary 
star. The very different depth of eclipses is actually  due to the combination of high eccentricity and orbit orientation in space. 

It was interesting to check, a posteriori, that this single light curve solution gave the following deviations for 
the various parameters with respect to the combined solution of photometry and radial velocity data: 6\% for $e$ 
and $\omega$, 1\% for $i$, 15\% for $T_{\mathrm{eff,2}}$, 6-15\%  for the component radii 
($r{_\mathrm{1}}$,$r{_\mathrm{2}}$), 12\% on $L_{1}$. We did not adjust the mass ratio, and actually the true value 
is outside the range we explored, but we had indeed a tendency of smaller values of the cost function  for the 
higher $q$ values. Furthermore, the value of $F_1$ was within 7\% of that derived  later on from line profile fitting. On the other hand the value of $F_2$ was almost 4 times the spectroscopic one. 

In conclusion the light curve solution was  well constrained for the main parameters and for primary rotational 
distortion.  This has to be put in relation to the high eccentricity signatures which act as pinpoints
on the parameter values.
\begin{table}
\caption{Radial velocities of HD 174884}       
\label{rvct}      
\centering                          
\begin{tabular}{c r r}        
\hline\hline                 
BJD -245000 &  $v_{\mathrm{rad,1}}$  &  $v_{\mathrm{rad,2}}$  \\    
                      &    (\ks)     & (\ks)   \\
\hline                        
4625.6910668  & -94.5 $\pm$1.6 &  85.1$\pm$2.6 \\
4625.7333949  & -97.1 $\pm$1.7&   87.6$\pm$2.6 \\
4625.7767647  & -97.5 $\pm$1.5&  99.5 $\pm$2.5 \\
4625.8241393  & -104.4$\pm$1.5 &  99.3$\pm$2.5 \\ 
4625.8549044  & -107.6$\pm$1.5 &  99.0$\pm$2.5 \\
4626.6823677  & -78.7 $\pm$1.6 &  69.8$\pm$2.7 \\ 
4626.7347772  & -74.8 $\pm$1.6 &  63.9$\pm$2.7 \\ 
4626.7602527  & -76.1 $\pm$1.6 &  60.4$\pm$2.7 \\ 
4626.7880894  & -74.4 $\pm$1.6 &  58.2$\pm$2.8 \\
4626.8353945  & -70.3 $\pm$1.6 &  53.4$\pm$2.8 \\
4627.6872211  &  24.9 $\pm$2.0 & -83.9$\pm$2.9 \\
4627.7463070  &  31.6 $\pm$2.0 & -90.4$\pm$2.8 \\
4627.7684857  &  35.1 $\pm$2.0 & -98.2$\pm$2.7 \\
4627.8089965  &  36.6 $\pm$2.0 &-107.4$\pm$2.5 \\
4627.8536741  &  43.6 $\pm$1.9 &-113.1$\pm$2.5 \\
4628.6470598  &  77.1 $\pm$1.7 &-162.4$\pm$2.5 \\
4628.6771883  &  66.7 $\pm$1.8 &-152.9$\pm$2.5 \\
4628.7144813  &  58.6 $\pm$1.8 &-137.3$\pm$2.5 \\
4628.8275062  &  23.3 $\pm$2.1 & -91.6$\pm$2.8 \\ 
4628.8613847  &  15.2 $\pm$2.2 & -80.6$\pm$2.9 \\
4629.6750711  & -108.0$\pm$1.5 & 105.0$\pm$2.5 \\ 
4629.7286030  & -106.9$\pm$1.5 & 108.1$\pm$2.5 \\ 
4629.7752019  & -106.6$\pm$1.5 & 107.4$\pm$2.5 \\ 
4630.6845270  & -54.8 $\pm$2.0&   30.7$\pm$3.5 \\
4630.7230468  & -48.7 $\pm$2.0 &  23.9$\pm$3.5 \\
4630.7712660  & -44.7 $\pm$2.0 &  18.1$\pm$3.5 \\
4630.8161634  & -42.1 $\pm$2.0 &  13.7$\pm$3.5 \\
4630.8495673  & -32.2 $\pm$2.0 &   5.1$\pm$3.5 \\
4631.7076510  &  71.8 $\pm$1.7 &-159.7$\pm$2.5 \\
4631.7415409  &  77.3 $\pm$1.6 &-175.2$\pm$2.5 \\
4631.7768777  &  83.5 $\pm$1.7 &-176.7$\pm$2.5 \\
4631.8041587  &  85.6 $\pm$1.7 &-182.6$\pm$2.5 \\
4631.8305253  &  89.7 $\pm$1.8 &-187.0$\pm$2.5 \\
4632.6861425  & -32.9 $\pm$2.0 &   3.2$\pm$3.5 \\ 
4632.7415956  & -51.7 $\pm$2.0 &  25.0$\pm$3.5 \\
4632.7934837  & -62.5 $\pm$2.0 &  45.0$\pm$3.0 \\ 
4632.8199197  & -72.8 $\pm$1.8 &  48.0$\pm$3.0 \\ 
4632.8472122  & -75.2 $\pm$1.6 &  53.9$\pm$3.0 \\ 
4633.7128864  & -98.8 $\pm$1.6 &  92.8$\pm$2.5 \\
4633.7401557  & -95.9 $\pm$1.6 &  94.8$\pm$2.5 \\
4633.7913378  & -94.1 $\pm$1.6 &  87.2$\pm$2.5 \\
4633.8186997  & -91.9 $\pm$1.6 &  81.4$\pm$2.5 \\
\hline                                   
\end{tabular}
\end{table}

\subsection{Spectroscopic follow-up}
\label{sp_fu}
  \begin{figure*}
   \centering
   \includegraphics[width=8.0cm]{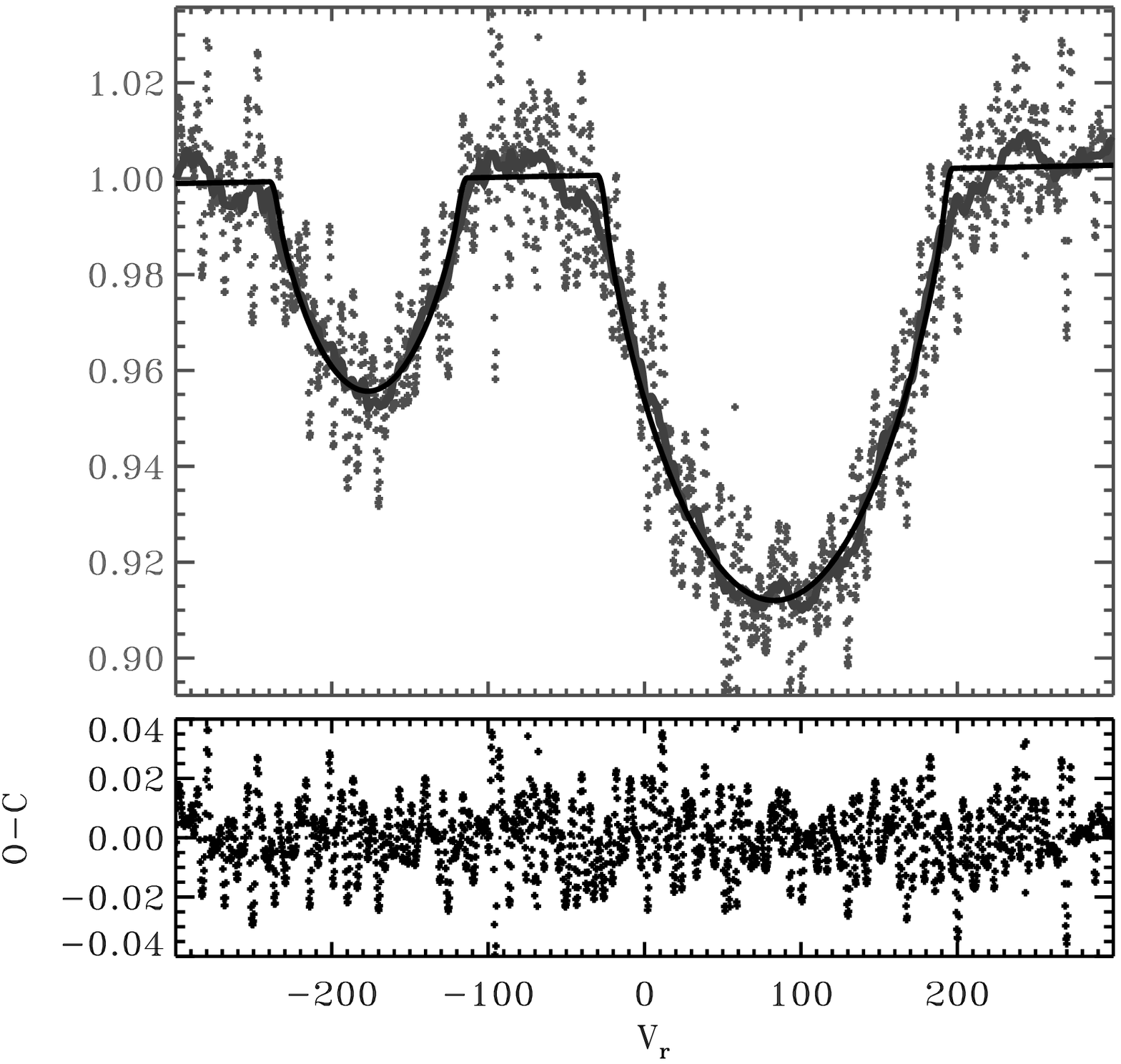}
   \includegraphics[width=8.0cm]{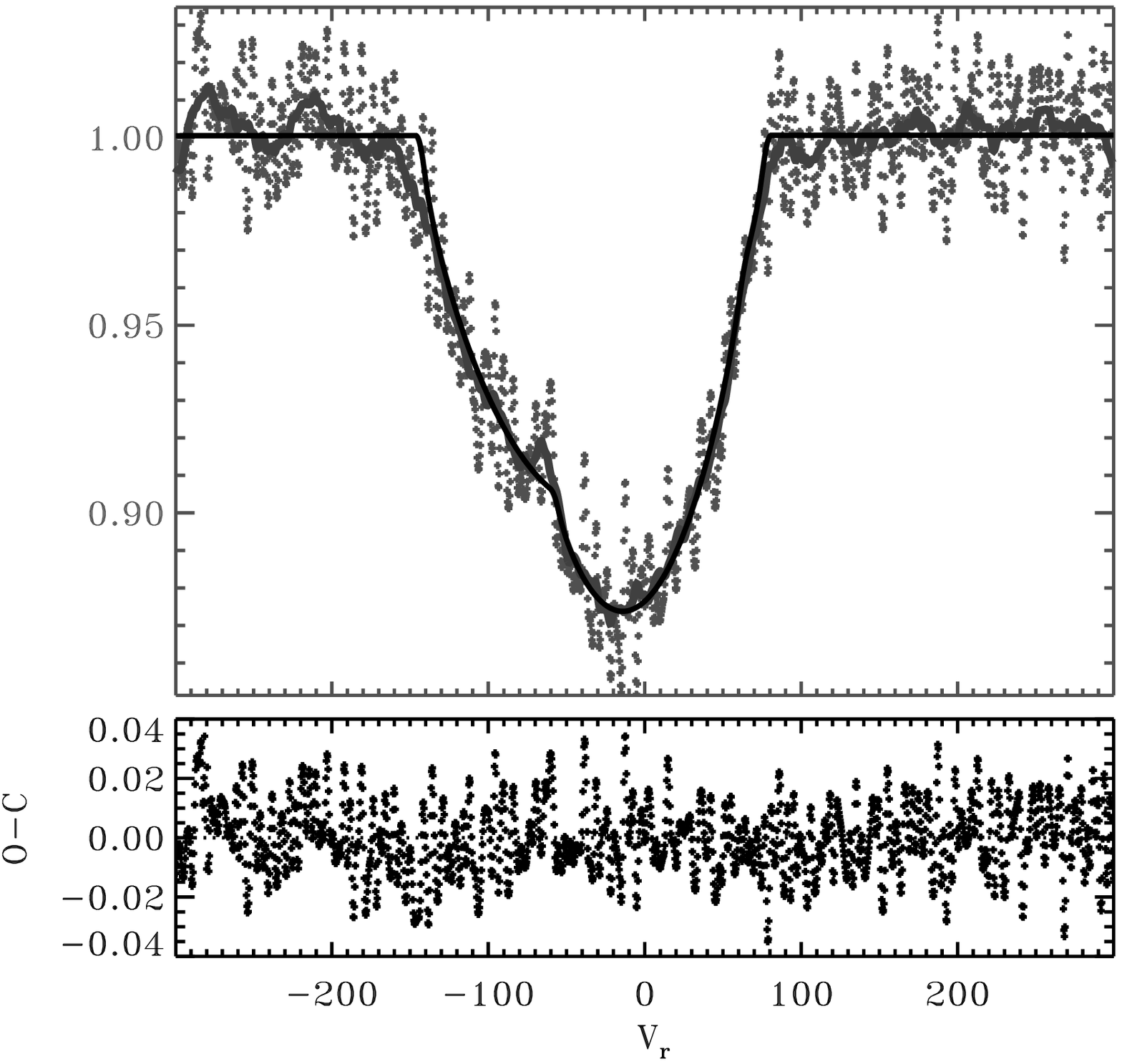}
      \caption{An example of the fit of the line profiles of  \ion{Mg}{ii}~4481~\AA\ .  The black solid line is the
fit with a function obtained
      by summing the individual profiles of the components  (a convolution of a gaussian and rotational
broadening). For clarity 
      a gray line is also shown, which corresponds to a smoothing with an averaging  boxcar of width comparable to
the FWHM 
      of the interstellar \ion{Na}{ii} D line (the narrowest lines appearing in the spectra).
      Left panel: unblended lines at phase $\varphi=0.71$,  right panel:  blended lines at $\varphi=0.45$. The
lower boxes show the fit residuals. 
              }

     \label{prof}
   \end{figure*}

Given that HD 174884 turned out to be a close binary, a spectroscopic
campaign was organised as soon as possible after the CoRoT space photometry had
become available to us. We used the CORALIE \'echelle spectrograph attached to
the 1.2m Euler telescope at La Silla in Chile. The star was monitored from 7 to
16 June 2008,   adopting an exposure time of 30 minutes. We used the online
reduction package available for the CORALIE spectrograph and based on
the method by Baranne et al.\ (1996). The process involves the usual steps of
de-biasing, flat-fielding, background subtraction and wavelength calibration by means of
measurements of a ThAr calibration lamp.  All spectra were subsequently
normalized to the continuum by fitting a cubic spline  and barycentric corrections 
were computed for the times of mid-exposure.  This led to barycentric 
wavelength-calibrated normalized spectra with a S/N ratio between 50 and 120 in the blue 
spectral region (the red parts suffered more from relatively poor atmospheric conditions 
during some nights).
 nspection of the spectra revealed that HD~174884 is a double-lined spectroscopic binary, evident
line doubling  could be seen in \ion{He}{i} and, more clearly in \ion{Mg}{ii}~4481~\AA\  line (see Fig. \ref{prof}), 
identifying both components as stars of spectral type B. The finding of similar temperatures from
the preliminary light curve solution was therefore fully confirmed.
 
The  set of 45 spectra was independently analyzed with different methods: profile fitting of the 
abovementioned \ion{Mg}{ii} line and Fourier spectra disentangling. The radial velocity curves from
profile fitting were used in the simultaneous solution discussed in  Section \ref{lc_finfit}.  
Fourier disentangling provided the individual stellar spectra  and  allowed, as well, a  cross-check with the results of
radial velocity analysis, that turned out to be  very useful, given the peculiarity of the 
system configuration.

  \begin{figure}[!hbt]
   \centering
   \includegraphics[width=8.7cm]{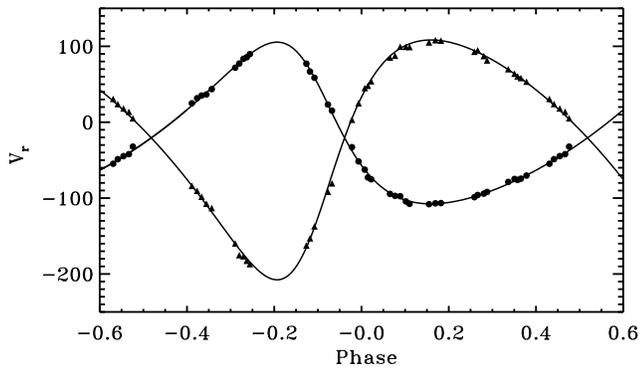}
      \caption{The radial velocity curves for the two components of HD~174884 (dots: primary , triangles: secondary component). The phases are computed according
      to the ephemeris in Eq. \ref{ephe}, velocity unit is kms$^{-1}$.   The solid line is the fit described in Section
	  \ref{lc_finfit}. 
              } 
     \label{rvc}
   \end{figure}
   
\subsubsection{Radial velocity curves from profile fitting} \label{RV_modeling}
We extracted the region 4450-4500 \AA\  and fitted the  \ion{Mg}{ii}~4481~\AA\ line profile with   a standard non-linear 
least squares fit procedure (IDL's CURVEFIT) using a user-supplied non-linear function with an arbitrary number of parameters. 
The supplied function was a theoretical profile obtained   by convolving for each star a Gaussian 
(corresponding to CORALIE spectra resolution) and a (dominant) rotationally broadened  profile. In principle there 
were 8 adjustable parameters (position, width, depth of each profile and the values of continuum on the two sides). 
However,  the iterative procedure did not converge in the case of heavily blended profiles.   
Therefore, we  fitted first  the  unblended profiles, adjusting the full set of 8 parameters,  and fixed then
the rotational
velocities, $v_{\mathrm{r}} \sin i$, to the average  values obtained in this way  for the  remaining fits. 
These average  values are  $v_{\mathrm{r,1}} \sin i= 109.3 \pm 1.5 $ kms$^{-1}$ and 
 $v_{\mathrm{r,2}} \sin i= 59.7 \pm 2.7 $   kms$^{-1}$. 

The fits were performed  on the original profiles and as well on the smoothed ones, with different boxcars 
(see Fig.~\ref{prof} for examples). The results were very similar even with resolution degraded down to 0.025 \AA/pix, i.e. the FWHM of the narrow NaI D interstellar lines (the narrowest lines seen in the spectra). The values in Table \ref{rvct} correspond  to the fit of the unsmoothed profiles. The fit uncertainties were derived with a bootstrap experiment (see Section \ref{bootstrap}), providing a more meaningful statistical estimate than formal errors of the fit.

\begin{figure*}[!ht]
\centering
\includegraphics[angle=-180,width=17cm]{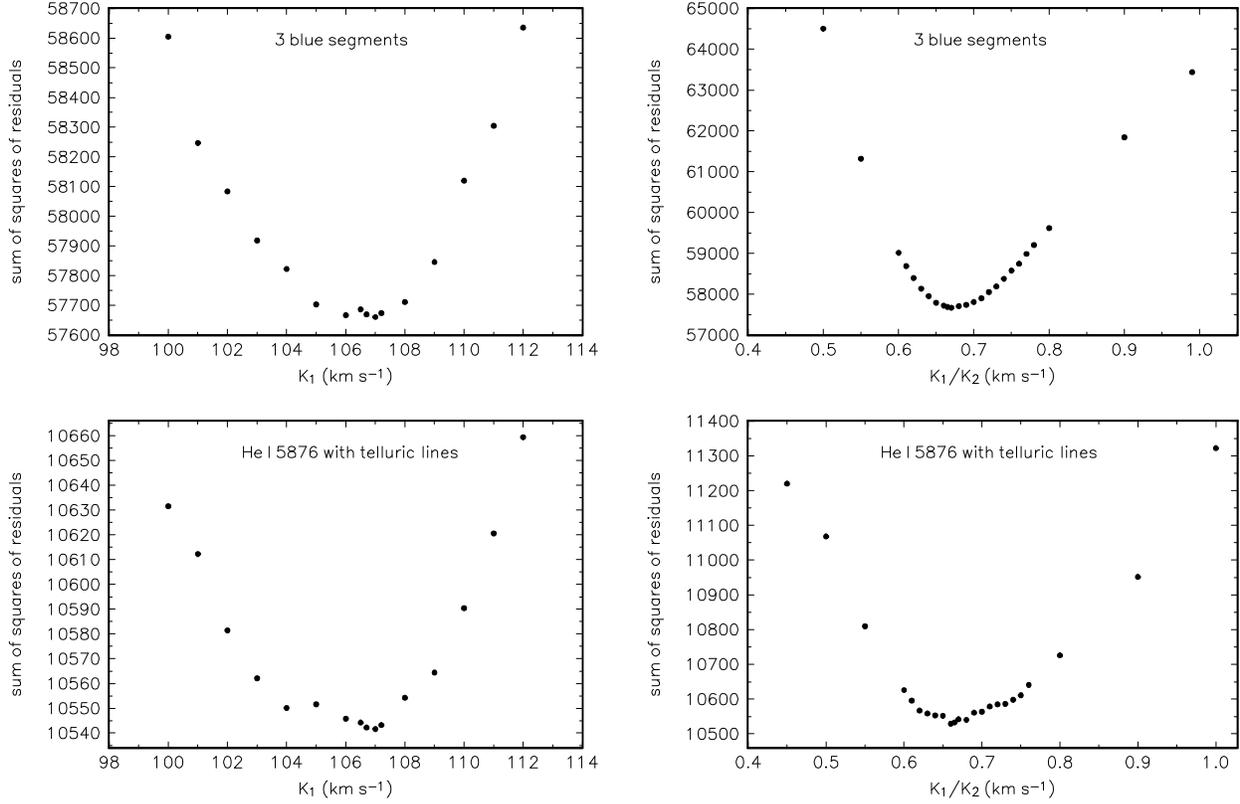}
\caption{A search for optimal values of $K_1$ and the mass ratio $K_1/K_2$.}
\label{kmap}
\end{figure*} 
Only 42 out of 45 spectra yielded a determination of radial velocities, for the remaining three, too close to conjunctions, the fitting procedure did not converge. The resulting values are collected in Table \ref{rvct} and the radial velocity curves are shown in Fig.~\ref{rvc}.
The radial velocity curves were fitted  with PHOEBE. This preliminary solution was not simultaneous, but we used
an iterative procedure between light and radial velocity curve solutions, feeding the output  of the iteration on one set of data as  input to the other. The purpose was to get a set of input parameters for the final combined solution
presented in Section \ref{fin_anal}. Actually the parameters obtained from the  preliminary light curve fit of Section \ref{prel_lc_anal}   were found to predict fairly well the radial velocity curves, apart of course from the mass ratio.
The values best fitting the radial velocity curves alone were: $e=0.292 \pm 0.003$, $\omega=51.85^{\circ} \pm 0.5$, $a \sin i = 18.27 \pm 0.08$~\rs, $\gamma=-20.7\pm 0.3$ kms$^{-1}$, $q=0.674 \pm 0.005$ (formal errors). The corresponding radial velocity amplitude for the primary  is $K_1=106.4 \pm 0.6$ \ks. 


\subsubsection{Fourier disentangling of spectra} \label{disentangling}

\citet{southworth2007} carefully compared various possible present-day
techniques to derive the orbital elements of spectroscopic binaries and
concluded that the best results are obtained via properly applied
disentangling. They showed that it is necessary to map the dependence
of the sum of squares of residuals in the semiamplitudes $K_1$ and
$K_2$ (or the mass ratio $K_1/K_2$) since this sum usually has a number
of local minima in the parameter space.

To obtain reliable orbital elements of \hd based on all 45 spectra we applied
the disentangling in the Fourier domain developed by
\cite{korel1,korel2,korel3}, using the latest publicly available version of
his Fortran program KOREL, dated December 2, 2004.

In general, our experience with KOREL fully confirms the findings
by \citet{southworth2007} and also that the result is sensitive to the choice
of starting values of the elements and initial values of the simplex
steps, i.e. that the simplex solution can easily be trapped in a number
of various local minima.
\begin{table}[!hb]
\caption[]{KOREL disentangling orbital solutions for the three combined blue
spectral regions near some stronger spectral lines useful for determination
of atmospheric parameters, and for the red spectral region near the
\ion{He}{I}~5876~\ANG\ line.  The period, eccentricity and longitude
of periastron were kept fixed at the values of 3\fd65705, 0.29394, and
51.$\!\!^\circ$31, respectively. The epoch of the periastron passage
$T_{\rm periast.}$ is tabulated in HJD$-$2454628.}\label{korel}
\begin{flushleft}
\begin{tabular}{rrrrrrlll}
\hline\hline\noalign{\smallskip}
Element&3 blue regions&\ion{He}{i} 5876\\
\noalign{\smallskip}\hline\noalign{\smallskip}
$T_{\rm periast.}$ &0.69173&0.69037\\
$K_1$(\ks)         &106.983&106.956\\
$K_2$(\ks)         &159.39&161.64\\
$K_1/K_2$          &0.6712&0.6617\\
\noalign{\smallskip}\hline
\end{tabular}
\end{flushleft}
\end{table}

 The spectra at our disposal have a large resolution 
and it would be
very time-consuming to carry out disentangling for the whole interval
of wavelengths they cover. At the same time, the $S/N$ of the spectra
is not particularly high. To cope at least partly with this problem
and to obtain good disentangled spectra containing spectral lines
sensitive to the determination of atmospheric paramaters and chemical
abundances, we slightly degraded the spectral resolution, averaging
the spectra over 0.05~\ANG\ and used only the following four spectral regions:
4050 -- 4190, 4450 -- 4500, 4810 -- 4910, and 5860 -- 5885 \ANG.
\noindent For each spectral segment, we estimated the $S/N$ ratio of
individual spectrograms and used the weights proportional to  $(S/N)^2$.
The disentangling was then carried out separately for the three blue
spectral regions combined into one solution and for the red region near
the \ion{He}{I}~5786~\ANG\ line where we also considered the telluric lines
in the solution. In all cases, we derived also the relative line strengths
for individual spectra during the solutions \citep{korel2}.
In all solutions, we kept the orbital period $P$, eccentricity $e$ and
the longitude of periastron $\omega$ fixed at the values of
3\fd65705, 0.29394, and 51.$\!\!^\circ$31, respectively, which were
derived from the final light curve fit (see next section). 
The epoch of the periastron passage
$T_{\rm periast.}$ was chosen inside the time interval covered by the
spectra, near HJD$-$2454628.7, but was allowed to converge.

\begin{figure}
\centering
\resizebox{8cm}{!}{\includegraphics{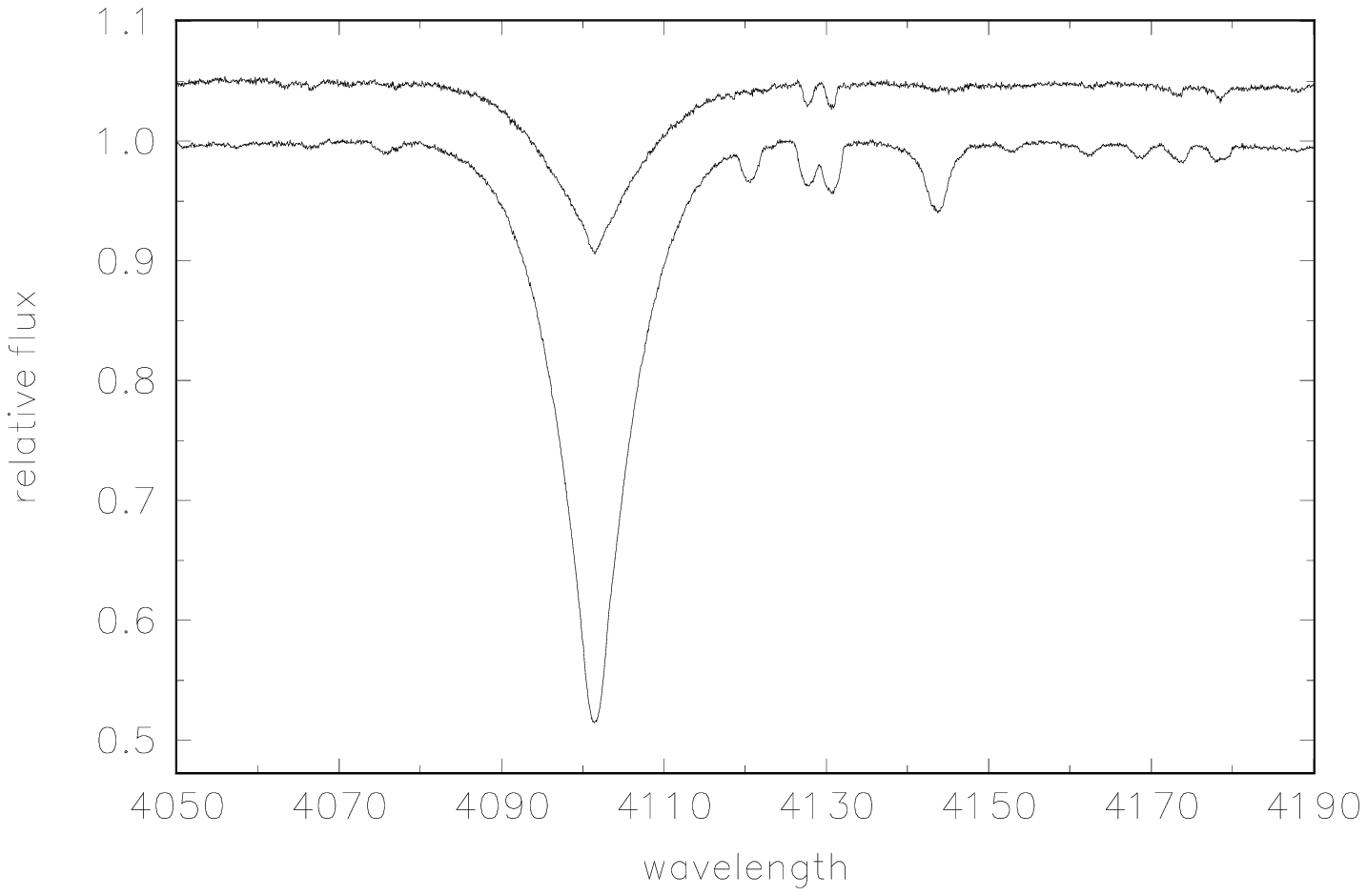}}
\resizebox{8cm}{!}{\includegraphics{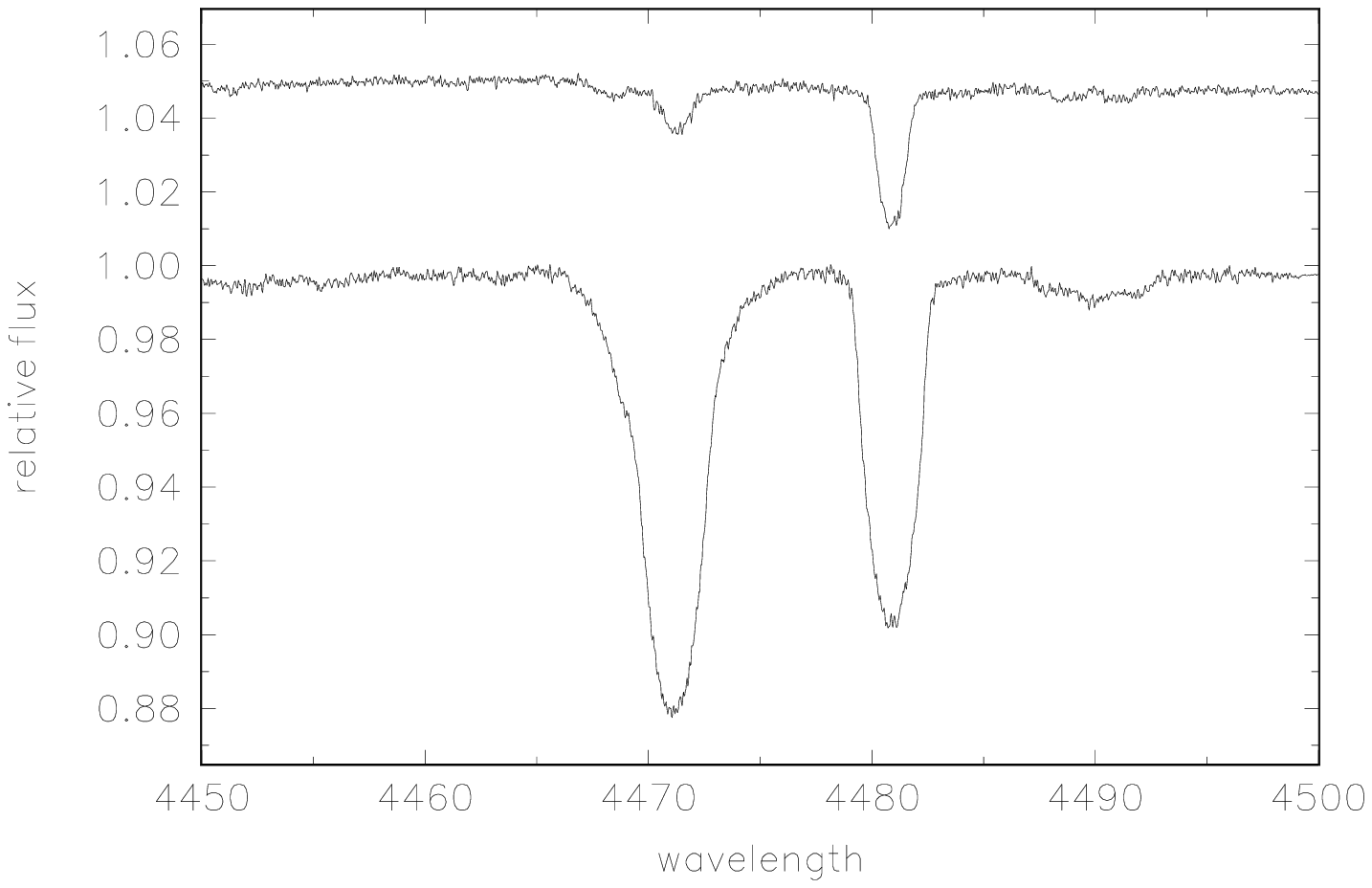}}
\resizebox{8cm}{!}{\includegraphics{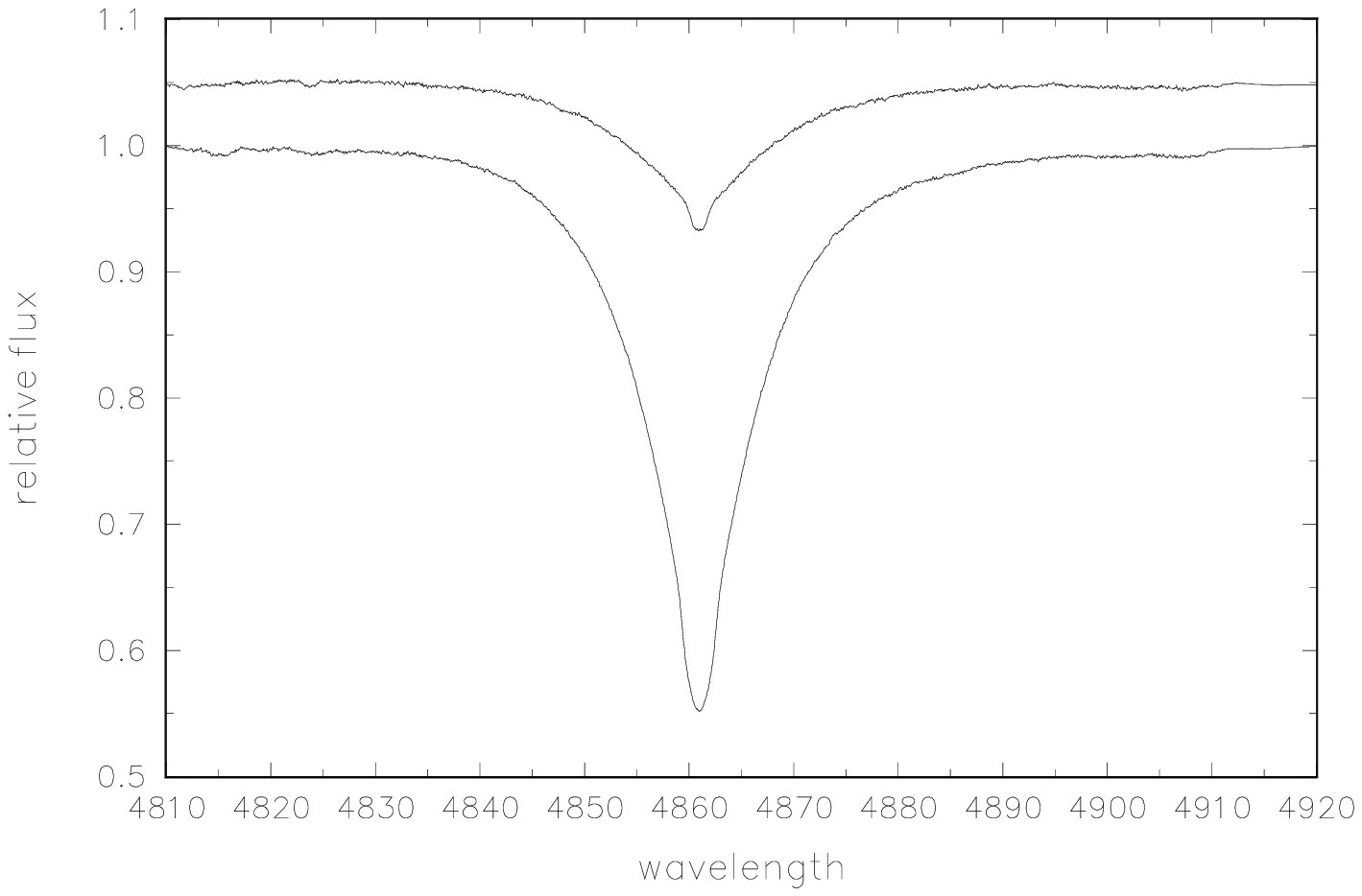}}
\resizebox{8cm}{!}{\includegraphics{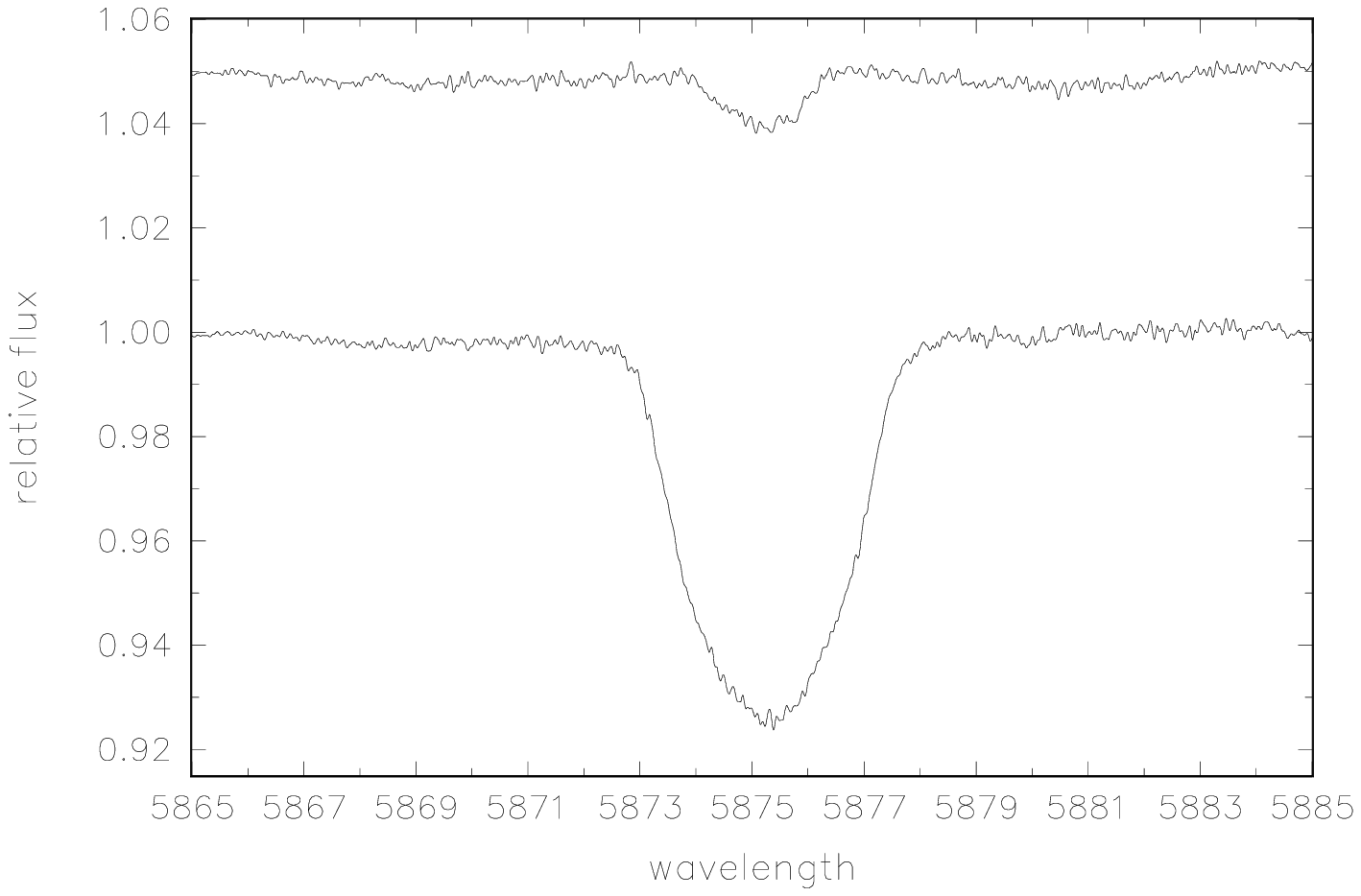}}
\caption{Disentangled line profiles.}
\label{profs}
\end{figure}
We first investigated the variation of the sum of squares of the residuals
as a function of the radial velocity semi-amplitude  of the primary only, keeping
all other elements but the periastron passage fixed. This dependence
was investigated separately for the combined blue, and the red spectral
segment and is shown in a graphical form in the left panels of Fig.~\ref{kmap}.
This led to the conclusion that the true semi-amplitude of the primary must be
close to 107~\ks. Fixing this value of the semi-amplitude, we then
investigated the sum of squares of residuals as a function of the mass ratio,
again separately for the blue and red spectral segments. This also rather
uniquely restricts the plausible values of the mass ratio to
0.66-0.67 (cf. the right panels of Fig.~\ref{kmap}).

The final disentangling was then derived starting with the best values
found and allowing for the convergence of $K_1$, $K_1/K_2$, and
$T_{\rm periast.}$. We did a number of trial solutions for each spectral
segment, varying slightly the initial values of the elements and
starting simplex steps to find out solutions with the smallest sums
of residuals. The results are summarized in
Table~\ref{korel}. Their comparison provides some idea about the probable
errors of the elements. We remind that for the region $5860-5885$~\ANG,
we also included the weak telluric lines into the solution.
The comparison of results is useful since the program KOREL we used
does not provide any estimate of the errors of the elements.

We included KOREL results in PHOEBE as cross-check, following a somewhat unusual procedure. For each of the four KOREL solutions we extracted radial velocities of both components for the 45 times of observations and derived their average
values over the two solutions for each BJD, giving 3 times higher weights to the blue RVs based on the separate spectral segments. These ``mean theoretical" RVs  were used to verify the results from PHOEBE that are based on profile fitting RV curves; we find excellent agreement between the two methods of RV extraction.

Keeping the elements from the final PHOEBE solution, discussed in the next
section, fixed, we disentangled the binary spectra. They are shown in
Fig.~\ref{profs}.

\section{Final analysis} 
\label{fin_anal}

\subsection{Simultaneous light and radial velocity curve fit}   
\label{lc_finfit}
To obtain a set of parameters that most accurately describes physical and geometric properties of HD~174884, we
submitted photometric and RV data in the time domain to a simultaneous fitting process in PHOEBE. Simultaneous fitting
imposes consistency of the solution across all available data-sets.

The initial values of $T_{\mathrm{eff},1}$ and $T_{\mathrm{eff},2}$ were adopted from the previous preliminary fits:
$13,140$\,K and $11,000$\,K, respectively. The values of gravity darkening coefficients and albedoes were  set to
their theoretically expected values for radiative envelopes: $g_{1,2} = 1.0$; $A_{1,2} = 1.0$.  The center-of-mass radial velocity
$v_\gamma$ was derived from the RV fit and held constant.

We used the RV derived value of $a \sin i = 18.27 \pm 0.08$~\rs (cf.~\S\ref{RV_modeling}) to constrain the solution to those values of $a$ and $i$ that preserve $a \sin i$. Line profile analysis and spectrum disentangling
(cf.~\S\ref{disentangling}) yield consistent values of $v_1 \sin i = 109.3 \pm 1.5\,\kms$ and $v_2 \sin i = 59.7 \pm 2.3\,\kms$; these were used to constrain synchronicity parameters $F_{1,2}= \omegaup_\mathrm{rot}/ \omegaup_\mathrm{orb}$, i.e. the ratio of rotation to mean orbital angular velocity. Square-root limb darkening coefficients  are interpolated on-the-fly from PHOEBE lookup tables, based on the current values of $T_\mathrm{eff}$, $\log g$, $[M/H]$ and $v_\mathrm{rot}$ for each star. As before, passband luminosity was computed rather than adjusted, to enhance convergence. The constraints are imposed during each iteration, projecting out only those solutions that satisfy the embedded physical requirements.

Parameters that were adjusted are: eccentricity $e$ and argument of periastron $\omega$, inclination $i$, surface potentials $\Omega_1$ and $\Omega_2$, mass ratio $q$, and secondary surface temperature $T_{\mathrm{eff},2}$. The mass ratio typically has little detectable signature in well detached EB light curves since it affects ellipsoidal variations; in the case of HD~174884, however, the  CoRoT light curve's sensitivity on $q$ is significant, due to the eccentric orbit, the consequent star shape distortion
and high data accuracy.  These factors allowed us to adjust the value of $q$ based on simultaneous LC and RV data.
There is an excellent agreement between  the $q$ value from the final fit and that derived from KOREL.
The difference is less than  1-sigma interval   for the three blue segments and within 2 sigma for the red one (which has  a lower S/N ratio). 
 
The final  parameter set is listed in Table \ref{eb_params}.
     
\begin{table}
\caption{Physical and geometric parameters of HD~174884.}
\label{eb_params}
\centering
\begin{tabular}{lccc}
\hline \hline
Parameter:                             &                      & System             &                      \\
                                               & Primary              &                    & Secondary            \\
\hline
$a$ (\rs)                     &                      & $18.88 \pm 0.14$ &                      \\
$i$ (${}^\circ$)            	      &                      & $75.35 \pm 0.03$     &                      \\
$q=\mathscr M_2/\mathscr M_1$     &                      & $0.674 \pm 0.009$  &                      \\
$e$                                         &                      & $0.2939 \pm 0.0005$  &                      \\
$\omega$                               &                      & $51.31^o \pm 0.06^o$ &                      \\
$v_\gamma$  (\ks)                  &                      & $-20.6 \pm 0.3$    &                      \\
$\Omega$                              & $6.055 \pm 0.008$   &                    & $7.866 \pm 0.020$    \\
$T_\mathrm{eff}$ (K)              & 13140  &                 & $12044 \pm 50$      \\

$\mathscr M$ (\ms) & $4.04 \pm 0.11$      &                    & $2.72 \pm 0.11$      \\
$\mathscr R$ (\rs) & $3.77 \pm 0.03$    &                    & $2.04 \pm 0.02$    \\
$M_\mathrm{bol}$                  & $-1.71 \pm 0.04$   &                    & $0.01 \pm 0.05$   \\
$F = \omegaup_\mathrm{rot}/
         \omegaup_\mathrm{orb}$   & $2.0^* \pm 0.2^*$ &                  & $2.3^* \pm 0.2^*$ \\
\hline
\multicolumn{4}{l}{\rule{0pt}{2.6ex} ${}^*$computed from the observed $v \sin i$ and orbital period.} \\
\end{tabular}
\end{table}

Before adopting our final solution and proceding to comparison with the stellar models we felt there were still
 two  questions to address:  solution uniqueness and a realistic estimate of the uncertainties. 

It is well known that light curve solutions can be affected by uniqueness problems, and  this could be a serious issue in the case of our light-curve that features a grazing eclipse. It is also well known that the formal errors derived  from least squares covariance matrix are inaccurate because the parameter space axes are not orthogonal and marginalization around the solution is not fully justified.
More realistic errors were  derived   by bootstrap and  by   full-fledged heuristic scanning, performed 
 on all adjusted parameters. The latter procedure  provides information  on the morphology of 
 the cost function hyper-surface in the adjusted parameter space.
\begin{figure}[!ht]
   \centering
    \includegraphics[width=8.7cm]{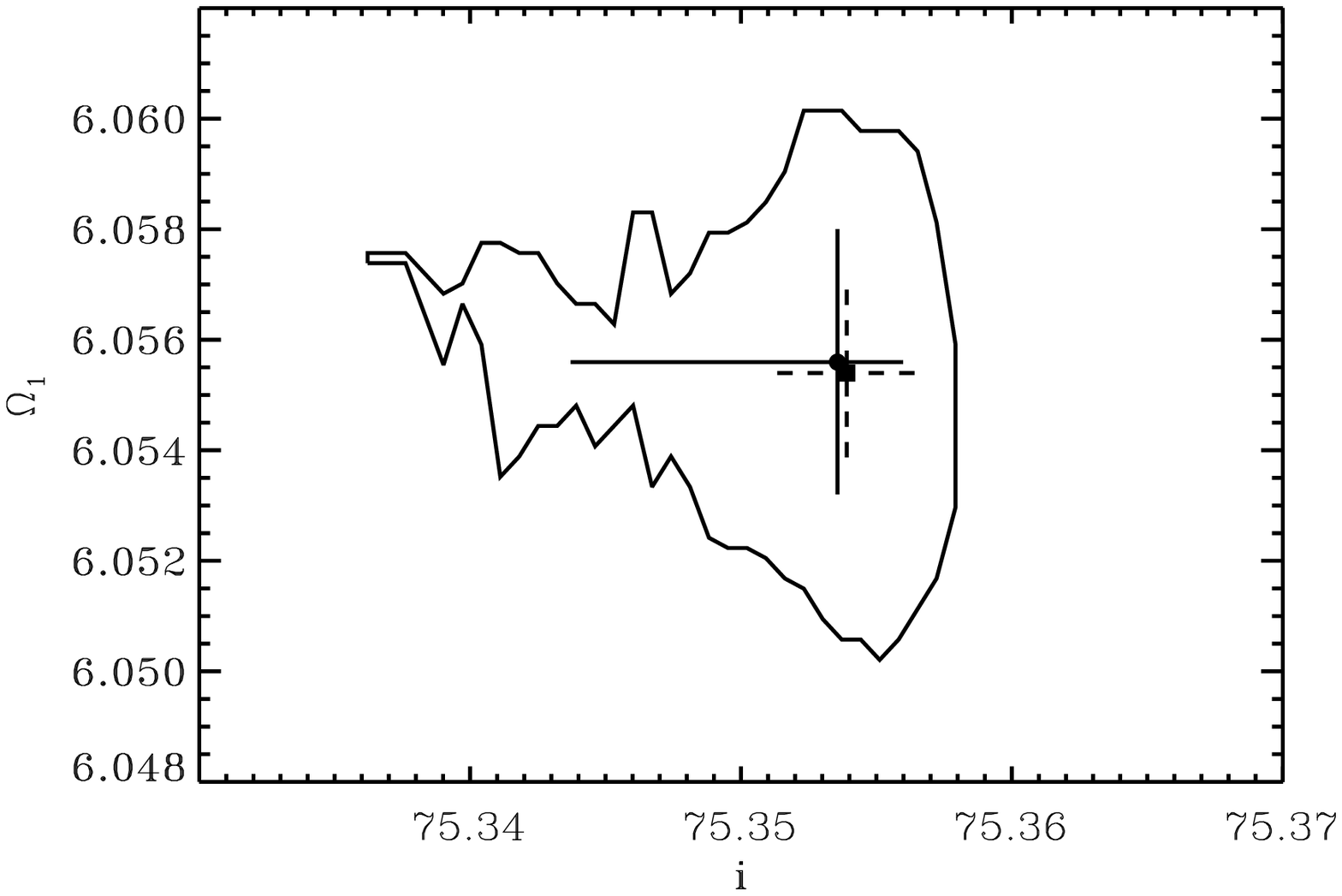}
    \includegraphics[width=8.7cm]{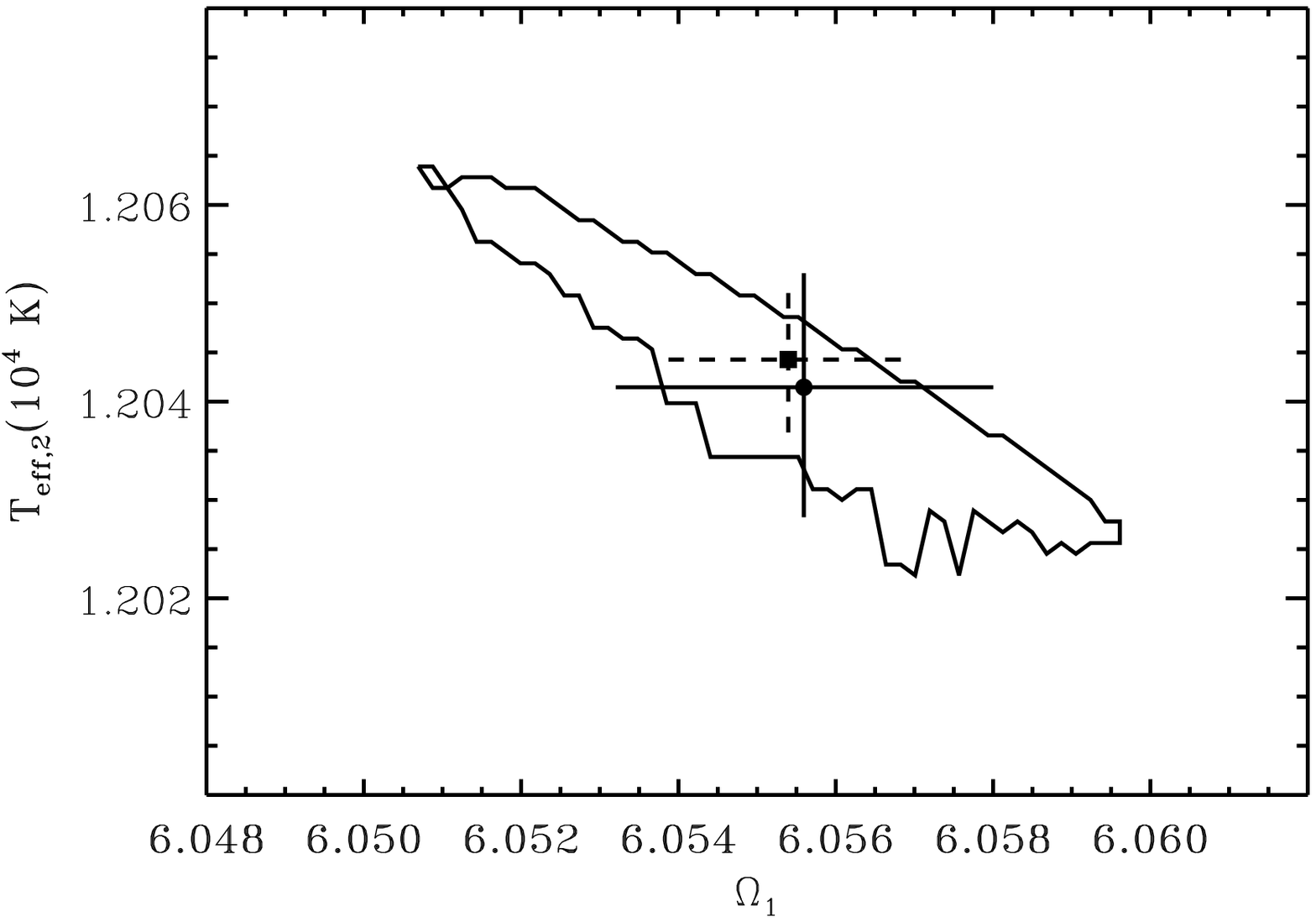}
     \caption{The contour  enclosing 68.3\% of all bootstrap solutions (1-sigma 
uncertainties) for two pairs of parameters: upper panel $i$ -- $ \Omega_1$, lower panel  $ \Omega_1$ -- $T_{{\rm eff,2}}$. 
 The solid  cross is the  median of  bootstrap solution values; the arm length is equal to  the uncertainty on the single parameters, 
not taking correlation into account (i.e the interval of the parameter containing 68.3 \% of solutions). The broken-line cross is our final solution 
with arms equal to the standard errors.
The small offset is due to the fact that bootstrap solutions, because of repetitions, are obtained from a curve with a different (and worse) 
sampling than the original light-curve.
The  projections of the contours on the axes provide the uncertainties of  (correlated)  parameters mentioned in the text.}
      \label{boot}
     \end{figure}
 The final estimates of most parameter uncertainties  appearing in Table \ref{eb_params},  is  obtained by the heuristic
 scan presented in Section \ref{heuristic_scan}.
 As, however, this is a procedure rather expensive in terms of computing
time and resources , we found it instructive  to compare the results of both methods, and quantify the differences.
Furthermore, we adopted the uncertainties from bootstrap for the parameters not adjusted in the scan.

 \subsection{Parameter uncertainties by bootstrap} 
\label{bootstrap}

Bootstrap resampling is a very useful technique to estimate parameter confidence levels of least squares solutions 
\citep[see, for instance][]{NR}. The method consists in generating many different data samples by random resampling 
with repetitions (bootstrapping) the available data, performing the minimization procedure for each sample and deriving
confidence intervals from  the resulting distribution of parameters. The advantage, with respect to the errors 
derived from least squares solutions, is that  a plot of the parameter distributions  directly shows the effect of 
inter-parametric correlations, and that the confidence intervals are not linked to Gaussian distributions of the residuals. 
In the parameter space   the one-sigma confidence levels, for instance of a pair of correlated parameters, is 
defined as the projection on the parameter axes of the contour containing 68.3 \% of the bootstrap solutions.   

 In this paper we used bootstrap resampling  to estimate the  errors on radial velocities, as obtained  from  
 profile fit. This is a standard application of the technique: the fitting procedure is repeated 
 (1000 times), after  bootstrapping the  fit residuals. 
   In principle the same technique can be used also to estimate the uncertainties of 
 the light and radial velocity curve fit, with the advantage of deriving more  realistic
 errors  than that the  formal ones and of gaining insight on parameter correlations. 
 It is not straightforward, however, to apply the bootstrap technique to the complex light curve solution process, 
because a complete   minimization process should be performed  for thousands of bootstrap data-sets. A simpler approach 
was suggested by \citet{mr97} who applied bootstrap resampling within the minimum already established by a single 
iterated solution (that is, using only one set of  residuals and parameter derivatives).  
This approach, as  discussed in  detail in the abovementioned paper, provides  underestimated uncertainties with 
respect to a full bootstrap procedure, as it is applied already in a minimum, but nevertheless, the confidence 
level we estimated for the parameters  are always larger than the errors from least square minimizations and
have a clear statistical meaning.

The bootstrap uncertainty intervals  were, therefore, obtained repeating, in this case, 10000  times the last iteration of the fit, after  random
permutation  of the residuals.  Fig.~\ref{boot} shows two 2-D projections of the n-dimensional  distribution of the resulting
parameter sets ($i$ -- $ \Omega_1$ and $ \Omega_1$ -- $T_{{\rm eff,2}}$). The figure allows a direct comparison
between the standard deviations from the PHOEBE last iteration and two confidence intervals derived  from bootstrap:
 that of each single parameter, not taking correlation into account, i.e. its interval containing 68.3\% of solutions, 
 and the 
1-sigma confidence level  of parameter pairs (the contours), clearly showing  the effect of correlations
 (see the plot for the pair $ \Omega_1$ -- $T_{{\rm eff,2}}$).  The projection of the contours on the axes 
 provides  a measure of the uncertainty taking parameter correlations into account.

The error estimates obtained by this simplified bootstrap procedure are somewhat smaller but still the same order than those from the heuristic
scanning of next section. Taking the largest of the full width 1-sigma intervals  (as bootstrap errors are not symmetric) among various projections
and comparing with the uncertainties of the heuristic scan of Table \ref{eb_params} we have,  0.02 vs 0.06 for $i$ (bootstrap vs heuristic scan), 
0.01   vs. 0.016 for  $ \Omega_1$, 0.01 vs. 0.04 for  $ \Omega_2$ and, 40 vs 100 for  $T_{{\rm eff,2}}$. The formal errors of the fit are typically several times smaller. 

The upper panel of Fig.~\ref{boot} suggests as well a  complex shape of the cost function along  inclination,
as the 1-sigma contour corresponds to a bimodal distribution of the solutions.  That feature is  fully revealed by the heuristic scan.
 \begin{figure*}[!hbt]
\centering
\includegraphics[height=17.cm,angle=-90]{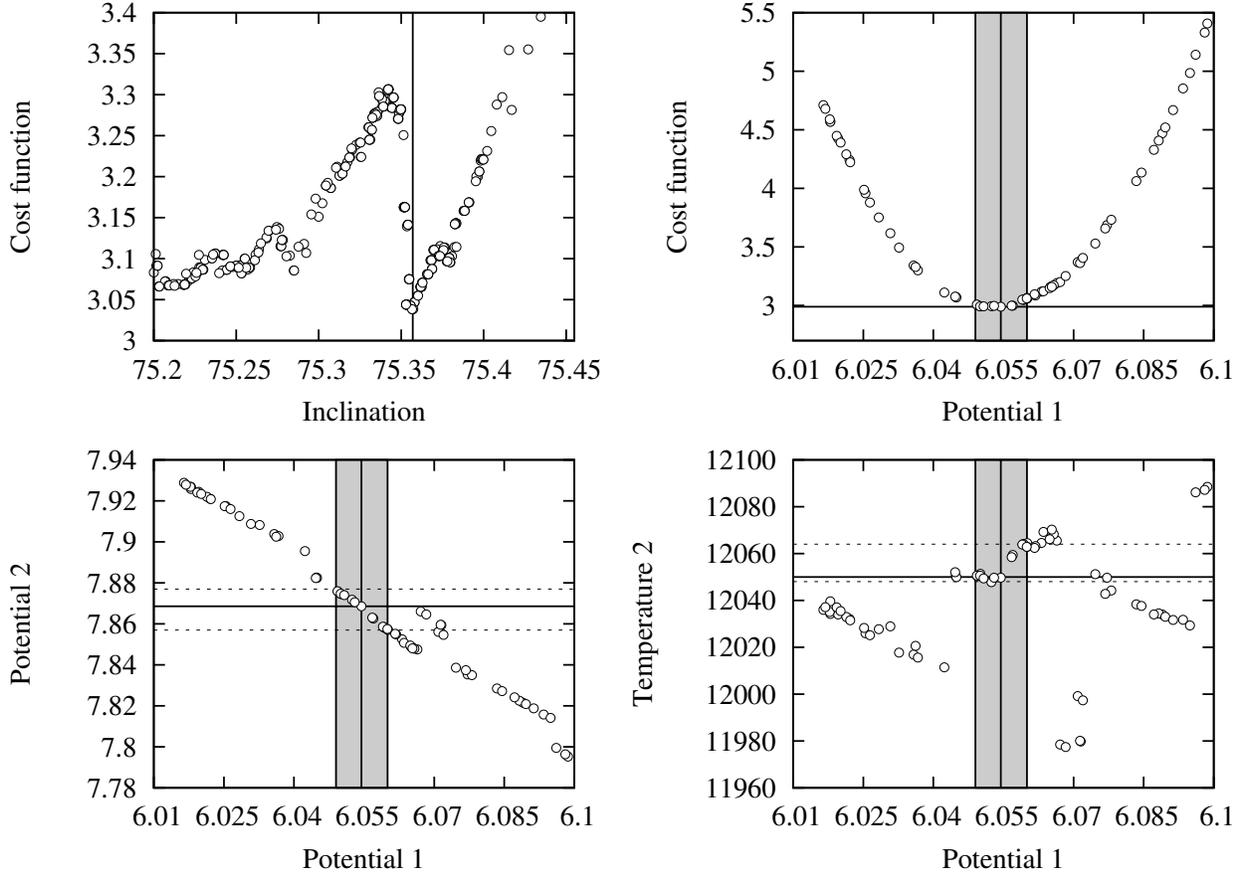}
\caption{The results of heuristic scanning. Top two panels depict the projection of the cost function space along inclination (left) and primary surface potential (right). The bottom two panels depict correlations between the secondary surface potential (left) and secondary effective temperature (right). A gray stripe encompasses 1\% deviations from the minimum cost function value in the primary surface potential, obtained by fitting a parabola to the top right panel. Solid lines represent the best-fit solution, while dashed lines represent parameter limits as determined by the plausible primary surface potential values.}
\label{heuristics}
\end{figure*}
 \subsection{Heuristic scanning of parameter space}
\label{heuristic_scan}

Heuristic scanning \citep{prsazw05} is a Monte-Carlo based method to map parameter correlations and cost function
degeneracy. Scanning produces an $N$-dimensional cost function map, where $N$ is the number of adjusted parameters. Projections in that $N$-dimensional parameter space reveal solution degeneracy.

We performed the scanning in two sequential steps: the first one along inclination, fitting $\Omega_1$, $\Omega_2$, $T_\mathrm{eff},2$ and computing $L_1$, $x_{LD}$ and $y_{LD}$, and the second one along the primary surface potential, fitting and computing the same parameters except for the inclination.

Because of the unprecedented accuracy of CoRoT light curve, a typical $\Omega_1$ degeneracy with inclination, surface potential $\Omega_2$ and effective temperature $T_{\mathrm{eff},2}$ is essentially broken. There is a clear $\chi^2$ minimum both in inclination (Fig.~\ref{heuristics}, top left panel) and primary surface potential (top right panel) that narrowly constrain the uniqueness of the solution. It was very instructive to observe the behavior of inclination with the mass ratio: the cost function profile varies quite substantially with the changing value of $q$. Although the location of the global minimum remains unaffected, adjacent local minima gain on depth, predominantly at the expense of ``sacrificing" the fit to the secondary minimum to better match the abundant out-of-eclipse regions. This indicates that the model we are using to fit the data is pushed to its limit and systematic errors begin to have a significant impact on the overall error budget of the solution.

\section{Physical properties of HD~174884}
\subsection{Comparison with evolutionary models}
\label{phys_par}

The comparison between the physical parameters obtained for HD~174884 components and the stellar models  was  done according to  the method described in \citet{mm05} and \citet{mmm07}. That consists in applying  a Levenberg-Marquardt gradient descent algorithm \citep{bev03} to minimize a quality function which describes the discrepancy between the models and the observables:
\begin{equation}
\chi^2=\sum_{i=0}^{No}\,\frac{(O_i^{obs}-O_i^{theo})^2}{(\sigma_i^{obs})^2}.
\end{equation}
The procedure iteratively adjusts free parameters of the models, yielding  $O_i^{theo}$, until best fit is reached.
In our case the ``observables" ($O_i^{obs}$)  are the system parameters derived in the previous sections: $\mathscr M_1$, $\mathscr R_1$, $T_{\rm eff,1}$, the effective temperature difference  between components, ($\Delta T_{\rm eff}$), the mass ratio ($q$), the radius ratio ($k$) and metallicity. The corresponding  errors are $\sigma_i^{obs}$. 
Though we initially chose as observables the absolute values (masses, radii, temperatures),  we finally preferred to
use  the   parameters of the primary component  and their ratios or differences with respect to the secondary, since the light curve solution is mainly sensitive to  relative values. For the uncertainty on $\Delta T_{\rm eff}$ we assumed, on the basis of the light curve fit results,  a value of  of 100~K.  As the available data were not accurate enough to derive metallicity,  we   assumed different values of $[{\rm M}/{\rm H}]$ with an uncertainty of 99\%.   \begin{table*}
\caption{Best fit models.  The values of the ``observables" , $O_i^{obs}$, are given for comparison in the note.}
\label{models}
\centering                          
\begin{tabular}{l c c c l l c c c c c c c }        
\hline\hline                 
 Model &$\mathscr M_1$ & $\mathscr M_2$ & $Z$ & $X$ & Age  & $OV_2$ & $\mathscr R_1$  & $\mathscr R_2$  & $T_{\rm eff\, 1}$   & $q$ & $k$ & $\Delta T_{\rm eff}$ \\    
            &  (\ms)& (\ms)& & &  (Myr) & &  (\rs)&(\rs)&(K) & & & (K)\\
\hline                        
 A         &  4.05$\pm$0.09 & 2.74$\pm$0.06 & 0.014$\pm$0.003 & 0.728$\pm$0.090 & 121$\pm$49& 0.000 & 3.76 & 2.03 & 13189 & 0.676 & 0.541 & 1631\\
 AC       &  4.04$\pm$0.11 & 2.75$\pm$0.08 & 0.011$\pm$0.003 & 0.720$^*$ & 112$\pm$9 & 0.062 & 3.68 & 1.97 & 13737 & 0.682 & 0.537 & 1520\\
 B1        &  3.99$\pm$0.08 & 2.74$\pm$0.06 & 0.011$\pm$0.002 & 0.740$^*$ & 131$\pm$9& 0.000 & 3.77 & 2.00 & 13097 &  0.688 & 0.530 & 1310\\
 B2        &  4.02$\pm$0.08 & 2.77$\pm$0.06 & 0.010$\pm$0.002 & 0.720$^*$ & 115$\pm$8 & 0.000 & 3.77 & 2.00 & 13614 &  0.689 & 0.530 & 1338\\
C6        &  4.01$\pm$0.08 & 2.76$\pm$0.06 & 0.012$\pm$0.002 & 0.720$^*$ & 118$\pm$ 9 & 0.200 & 3.77 & 2.00 & 13339 &  0.688 & 0.530 & 1295\\
 C1        &  4.05$\pm$0.11 & 2.78$\pm$0.08 & 0.018$\pm$0.010 & 0.738$\pm$10.2 & 131$\pm$57 & 0.654 & 3.77 & 2.02 & 12668 &  0.685 & 0.536 & 1305\\
 C4        &  4.04$\pm$0.11 & 2.77$\pm$0.08 & 0.015$\pm$0.007 & 0.740$^*$ & 130$\pm$13& 0.355 & 3.77 & 2.02 & 12870 & 0.686 & 0.535 & 1320\\
 C2        &  4.00$\pm$0.08 & 2.75$\pm$0.06 & 0.013$\pm$0.002 & 0.740$^*$ & 131$\pm$10& 0.200 & 3.77 & 2.00 & 12978 &  0.687 & 0.530 & 1277\\
 C5        &  4.00$\pm$0.08 & 2.75$\pm$0.06 & 0.012$\pm$0.002 & 0.740$^*$ & 131$\pm$1& 0.100 & 3.77 & 2.00 & 13045 & 0.688 & 0.530 & 1293\\
 C7        &  3.99$\pm$0.08 & 2.75$\pm$0.06 & 0.011$\pm$0.002 & 0.740$^*$ & 130$\pm$95& 0.050 & 3.77 & 2.00 & 13078 &  0.688 & 0.530 & 1298\\
\hline
\end{tabular}
\begin{list}{}{}
\item $O_i^{obs}$:  $\mathscr M_1=4.04\pm0.11$; $\mathscr R_1=3.77\pm0.03$;  $T_{\rm eff\, 1}=13140\pm1500$; $q=0.674\pm0.009$; $k=0.541\pm0.013$; $\Delta T_{\rm eff}=1096\pm100$ (or 1240$\pm100$, see text).
\item For all models $OV_1=0$. Model AC belongs both to group A and C. Several C-type models are listed to illustrate the effect of adjusting $OV_2$. 
\item  ${}^{*)}$ Fixed value.
\end{list}
\end{table*}

The theoretical  values ($O_i^{theo}$) are obtained from stellar evolution modeling with the code CLES
 \citep[Code Li\'{e}geois  d'\'{E}volution Stellaire, ][]{scuetal08}.
  In all model computations we used the mixing-length theory (MLT) of convection \citep{bv58} and the most recent  equation of state from OPAL \citep[OPAL05,][]{ronay02}. Opacity tables are those from OPAL \citep{iglrog96} for two different 
solar mixtures, the standard one from \citet[][GN93]{gn93} and the recently revised solar mixture from \citet[][AGS05]{ags05} 
. In the former $(Z/X)_\odot$=0.0245, in the latter $(Z/X)_\odot$=0.0167. These tables are extended at low temperatures
 with \citet{fergal05} opacity values for the corresponding metal mixtures. The surface boundary conditions are given by  \citet{kur98} atmosphere models.
     
 The parameters of the stellar model  are: mass,  initial hydrogen ($X_{\rm i}$) and metal ($Z_{\rm i}$) mass fractions, age and
 two  convection parameters ($\alpha_{\rm MLT}$ and the  overshooting  parameter $OV$).
  All of them, or just a subset, can be   adjusted in the minimization.
  The value of  $\alpha_{\rm MLT}$ was kept fixed, adopting  the solar value of 1.8, as it has no relevant effect on the
evolutionary tracks of models with masses in the domain of interest here.

  Different minimizations were done with  $X_{\rm i}$ as a free parameter or  fixed 
 to different  values between 0.69 and 0.74.  Similarly,   the overshooting parameter, which describes the size
 of the extra-mixed region close to the convective core, was kept fixed at values between 0.0 and 0.4, but in some cases  left free.
Binarity  puts an additional constraint on chemical composition and  age, which can be assumed to be the same for the components.  
Therefore, the number of free parameters  varied from four ($\mathscr M_1,\,\mathscr M_2,\, Z,\,$ age) to six : $\mathscr M_1,\,\mathscr M_2,\, X,\,Z,\, OV\,$ and  age. 
 We performed two sets of minimizations,  the first with the same physical description of components  
 (essentially same value of overshooting parameter), and the second allowing a different parametrization  of extra-mixing. 

\begin{figure}[!hb]
   \centering
    \includegraphics[width=8.7cm]{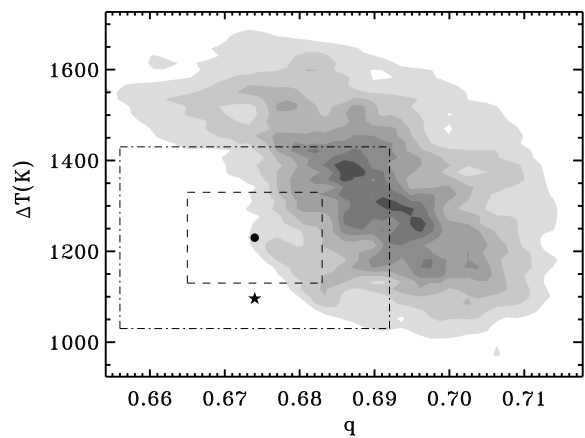}
     \caption{The iso-$\chi^2$ contours in the q--$\Delta T_{\rm eff}$ plane for the Levenberg-Marquard minimizations  of the deviations with respect to theoretical models.  Lower values correspond to darker colors. The plot clearly shows the correlation between the two parameters. The $\chi^2$ is computed with respect to the $\Delta T_{\rm obs}$ value obtained after correction for the effect of rotation (see text). The original value is also shown, as a starred dot. The broken line boxes  enclose  1- and 2-$\sigma$ intervals.}
     \label{chi2jm}
     \end{figure}

Regardless of these physical details  all computations show the same trend:  
if the $\Delta T_{\rm eff}$ is  well fitted (within 1-$\sigma$) the mass ratio is larger than the observationally derived value
and the deviation  is typically between   2 and 3$\sigma$ . This fact can be explained considering that  for two  stars with the same initial chemical composition and physical processes,   fixing the effective temperature difference at  given age is essentially equivalent to   fixing the  mass
difference.  Due to correlation, any combination of $\Delta T$ and $q$ yields estimates on $q$ that are inconsistent with the light and radial velocity curve solution. This behavior is evident in Fig.~\ref{chi2jm} which shows the iso-$\chi^2$ contours in the q--$\Delta T_{\rm eff}$ plane. This plot was built using the several thousands binary models computed in the minimization processes. If the constraint on $\Delta T_{\rm eff}$ is relaxed, the fit  within  1$\sigma$ for all the other constraints is achieved, but $\Delta  T_{\rm eff}$ is $ \sim \, 1600$~K  (Model A of Table~\ref{models}).  Among the  minimizations performed with same physics,  $\Delta T_{\rm eff}=1310$~K is the minimum difference of effective temperature that we were able to achieve,
 keeping the mass and radius ratios within 2$~\sigma$ (1.6$\sigma$ for $q$ and 1.3~$\sigma$ for $k$)  (Model B1).
These best fits were  obtained from computations without overshooting.  Values of $OV$ larger than zero  generally lead to  smaller
 primary masses and hence  larger $q$ values.

Is it possible to  reconcile the observed values of $q$ and $\Delta T_{\rm eff}$  with those from the theoretical models? 
To answer this question we considered the possible  alternatives and the weak points of comparison with the models. 
An  evident inconsistency between the stars  and the corresponding  models is that  CLES computes spherical non-rotating models,   
while the primary and  secondary component of HD174884 rotate, respectively,  at 25\% and 12\% of their break-up critical velocity, and 
are significantly non-spherical.
The  distortion of the primary  ($R_{pole}$ $\sim\, 2.5$\% smaller than $R_{\rm eq}$) is mainly due to  rotation. 
 It is well known that at given mass,  rotating stars appear to be cooler than non rotating ones, or -- expressed in different terms -- 
 at given effective temperature, non-rotating models will assign lower stellar masses than rotating ones. 
 The difference of effective temperature between rotating and non-rotating models depends on the rotational velocity,
 inclination of the rotation axis and chemical composition  of the star \citep[see, e.g., ][]{maemey04}. 
 An estimate of the average effect of rotation on the observed effective temperature for moderate rotators  is provided by \citet{bdm09} and their 
 results are in agreement with those  of the Geneva group for a 4 \ms\ star (Meynet, private communication). 
 According to the above-mentioned authors, the change in $T_{\rm eff}$ of a rotating star with respect to a non-rotating  one can be expressed as : 
 \begin{equation}
 \label{rotation}
  T_{\rm eff}(\eta)/T_{\rm eff}(0) = 1 - a \eta^2,
\end{equation}
 where  $\eta$ is the fraction of the critical break-up rotation rate, $T_{\rm eff}(0)$ is the effective temperature of the non-rotating model, and $a$ is a coefficient which changes from 0.17 at zero age to 0.19 at the end of the main sequence.
In the case of HD174884, the apparent decrease in temperature due to rotation amounts to   $\sim 160$~K for the primary  and  only  20~K
for the secondary component. As consequence, the observed effective temperature difference should be smaller than that of   non-rotating models,
 and that could explain the higher    $\Delta T_{\rm eff}$ values from best fit models with respect to that from the light curve solution. 
In this hypothesis, the constraint on  $\Delta T_{\rm eff}$ to be used with  our non-rotating models shall be increased from 1096~K to $\sim\,1240$~K. 
This value, however,  is still somewhat lower than that obtained in our best fit. 
 
 The value of the overshooting parameter, its dependence on the stellar mass, and the physical origin of the extra-mixing are still a matter of debate.
A  different value for the two components cannot be completely excluded and the hypothesis $OV_1= OV_2$ could be dropped.
  We tried, therefore, to satisfy our observational constraints by using fixed and different values of $OV_1$ and $OV_2$. We also performed computations fixing $OV$ for the primary and varying the $OV$  for the secondary component, and - as well -  keeping fixed $OV_2$ and deriving $OV_1$. The corresponding  models are
labeled as models C in Table~\ref{models}. An  overshooting parameter higher for the secondary than for the primary slightly improves the fit with respect to model B.  From a physical point of view, we could justify a secondary with a stronger mixing in the center, even if its rotational velocity
is lower than that of the companion, by invoking an  important gradient of the internal  rotational profile produced by  braking in the synchronization process. However,  the improvement in the fit is not  significant enough to  justify the introduction of an additional parameter.

The age of the system is similar for all models and has an average value of $125 \pm 7$~Myr.
 
  Finally, we can ask ourselves if our conclusions could depend on the evolutionary code we used for computing  stellar models.  
 A convincing answer to this question can be found in  the  extensive study which was carried out by the CoRoT/ESTA (Evolution and Seismic Tools Activity) group, in view of the CoRoT mission and the foreseen analysis of seismic data of the highest ever achieved accuracy.   The team 
made an accurate comparison among different evolutionary and pulsation codes for selected test cases. Their results \citep[][ and other contributions appearing in the same special issue]{lebmon08}   allow us to conclude that in the part of the parameter  space of interest here (MS stars of intermediate mass) stellar evolutionary codes adopting  the same physics  yield the same fundamental parameters and the same internal structure.
     
\subsection{Orbital evolution}
\begin{figure}
   \centering
    \includegraphics[width=8.7cm]{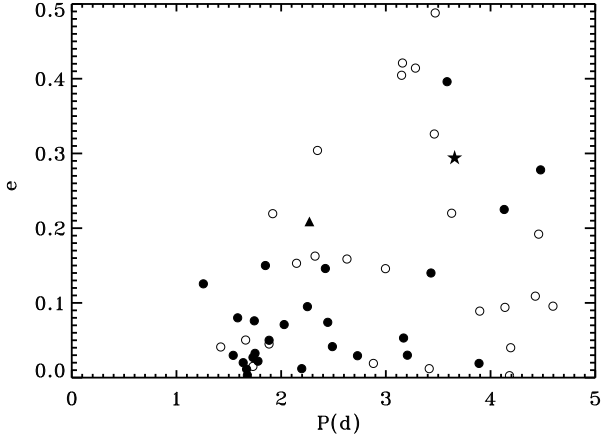}
     \caption{The period - eccentricity diagram, for close  binaries  according to the catalog of \citet{hege05}. Filled circles
     correspond to B~stars,  open circles to A-G stars.   
     The star symbol indicates the location of HD~174884, the triangle that of HD~313926 \citep[][see text]{smr07}.  
 }
     \label{pecc}
	 \end{figure}

The position of  HD~174884 in the period-eccentricity diagram of close binaries  is shown in Fig.~\ref{pecc}.  
The plot  is based on the  catalog of \citet{hege05}, with the addition of \object{HD 313926}, a  short period (P=2\fd27) high 
eccentricity (e=0.21) eclipsing binary formed by two B-type stars, which  was recently  discovered by the MOST satellite \citep{smr07}.
The  majority of B-binaries in the figure have main sequence components.
HD~174884 has the second highest eccentricity among binaries with B~type components.  

 \citet{smr07} suggest that the upper envelope for B~type stars in the period-eccentricity plot might be flatter than
 that of the later stellar types. This can be understood in terms of stronger dissipation, 
 and hence more efficient circularization mechanisms, for stars with larger fractional radii, see for instance
 the review of \citet{zahn05}.  That is at the origin of the overall trend in the plot (lower eccentricity for shorter periods),
  besides, at fixed period, less massive (unevolved) stars  have  smaller fractional radius and hence longer circularization timescales. 
 The evidence of the second effect is, however, marginal in our updated plot, as HD~174884 occupies  a region 
 previously empty of B-stars. 

 The relation between eccentricity and the fractional radius of early type binaries  has been extensively studied. 
 \citet{giuma84} analysed a  sample of galactic binary and   \citet{northz03}  more homogeneous ones in the Magellanic 
 Clouds.  All samples indicate the existence of a ``cut-off" fractional radius, above which all systems
have circular orbits. This value -- practically independent of chemical composition -- is $ \simeq 0.25$. The fractional
radius of HD~174884's primary is $r_{1} \simeq 0.2$, i.e. a  value consistent with its elliptic orbit.  
 Its position in the eccentricity vs. fractional radius plot of \citet{giuma84} (not shown here for the sake of brevity) 
 marks again  the upper envelope of the galactic sample for that period.
   
 The same mechanism producing orbit circularization is at the origin of spin-orbit synchronization on
 timescales typically a few orders of magnitude shorter than that for circularization, as -- in presence of
 efficient angular momentum transfer -- the ratio of the two timescales equals that  of  rotational 
 and orbital angular momentum content. 
As long as  the orbit is  elliptical, the component rotation will synchronize with the orbital motion at periastron.
 Since $v \sin i$ was  estimated from spectral analysis ($v_1 \sin i = 109\kms, v_2 \sin i=60\kms$) and the
eccentricity is well constrained from spectrophotometric modeling, we can readily verify whether this is the case.
 A theoretical value \citep{hut1981} of the synchronicity parameter for
synchronous rotation at periastron is $F=\sqrt{(1+e)/(1-e)^3}$,  and for our system $F=1.92$, while spectroscopic analysis yields
 $F_1 = 2.0 \pm 0.2 $ and $F_2 = 2.3 \pm 0.2$. Both stars thus seem to be  rotating marginally super-synchronously.
We can check if the dynamical state of the system is in agreement with the expected timescales.
If we follow Zahn's formalism \citep[see the above-mentioned review and][]{clacu97} to estimate the 
circularization and synchronization timescales of \hk,  adopting  a value of $\beta=0.2$ for the
fractional gyration radius \citep{clagim05} and $\log E_2=-6.6$ for the value of the second torque constant 
\citep{clacu97}, we get: $\tau_{\rm circ}= 8.9 \cdot 10^8$~yr ( or 35\% shorter considering the
contribution of both components), $\tau_{\rm sync,1} \simeq \tau_{\rm circ}/120 = 7.4 \cdot 10^6$ yr and 
$\tau_{\rm sync,2} \simeq 4 \tau_{\rm sync,1}$.
While the relations providing  these estimates are derived under the hypothesis of small deviation from circular orbit and
 synchronism, the resulting figures are indeed in good agreement with the dynamical status of
our system.

 The fact that in the same period interval of the $e$--$P$ plot both circularized and elliptic orbits can be
 found is interpreted \citep{zahn05} as due to much higher efficiency of tidal damping 
 when resonance locking  \citep{wittesav99} takes place. As this event is very sensitive to the binary parameters
 it might appear only in some  systems among those of similar period, depending on the other stellar parameters.
According to  this interpretation  the orbital evolution of HD~174884 has not  been driven by resonance locking.
\subsection{Pulsational properties: analysis of light-curve fit residuals}
\label{four_anal}
\begin{figure}
   \centering
     \includegraphics[width=8.7cm]{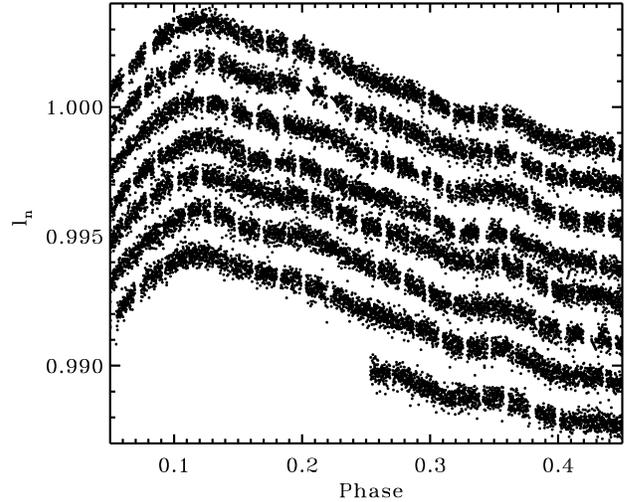} 
     \caption{Sections of the original light curve, phased according to the orbital period and 
	 vertically shifted for better clarity. The seven observed orbital periods go  bottom-up.
	 The oscillatory pattern of the residuals is clearly seen in each curve.
 }
     \label{lc_shift}
	 \end{figure}

The residuals from the phased light curve fit, appearing in the lower panel of Fig. 
\ref{phased_lc}, clearly show  a phase-locked pattern, which can be described in terms of two 
components: a complex irregular structure  at phases close to (and within)  primary 
eclipse and a multi-periodic oscillatory  behaviour at other phases. 
Their different nature  is suggested by the fact that only the latter
is  present in the original light curve  (see Fig.~\ref{lc_shift}).

We  interpret the first component as due to  systematic effects caused by the
 employed EB model, 
e.g. to small deviations in model atmospheres (and consequently in the limb darkening description),  
and   to the description of  stellar surfaces by a finite, though large, 
number of  elements.  These effects are expected to be  visible 
during eclipses, when one stellar disc eclipses the other. 
Those related to the  surface discretization are typically below 0.1 mmag, but  can 
be larger at some critical phases.
\begin{table*}
\begin{minipage}[t]{\columnwidth}
\caption{Results of Fourier analysis}       
\label{freq_tab}      
\centering                          
\renewcommand{\footnoterule}{}  
\begin{tabular}[ht!]{llllll}
\hline\hline                 
F$n$&Frequency  &Amplitude& Phase& S/N & Remark\\ 
& (d$^{-1}$) &  ($10^{-3})$ & \\
\hline
F1&	0.273444                    &    0.452    $\pm 0.002$\footnote{all uncertainties are the fit formal  errors from Period04}  &   0.3848 $\pm 0.0006$ & & $f_{\rm orb}$, fixed \\
F2&	0.5470  $\pm 0.0002$&	 0.284   $\pm 0.002$ &	0.840 $\pm 0.001$	&14.2 &  $\simeq 2 f_{\rm orb}$\\
F3&	2.1858 $\pm 0.0004$&	 0.120   $\pm 0.002$ &	0.349 $\pm 0.002 $	&12.0 & $\simeq 8 f_{\rm orb}$\\
F4&	3.5534 $\pm 0.0004$&	 0.111   $\pm 0.002$ &	0.822 $\pm 0.003 $	&11.2   & $\simeq 13 f_{\rm orb}$\\
F5&	1.0894 $\pm 0.0004$&	 0.097   $\pm 0.002$ &	0.840 $\pm 0.003 $	& 6.2 & $\simeq 4 f_{\rm orb}$\\
F6&	0.0631 $\pm 0.0004$&	 0.116   $\pm 0.002$ &	0.931 $\pm 0.003 $	&7.2   &   long term trend\\
F7&	0.8206 $\pm 0.0005$&	 0.091   $\pm 0.002$ &	0.570 $\pm 0.003 $	&7.8   & $\simeq 3 f_{\rm orb}$\\
\hline 
\end{tabular}
\end{minipage}
\end{table*}

 We attempted  to identify those segments of the code that need to be
restated in order to increase the model accuracy.  An important one is  the
above-mentioned surface discretization, which currently relies on equidistant
partitioning of the surface along stellar latitude and longitude. This
introduces a "seam" along $\varphi=0$ where all surface elements are
perfectly aligned. We have been devising a modified approach where
surface elements are determined by equidistant partitioning along the
length of the equipotential and shifted by half the element width at
$\varphi=0$. Preliminary tests indicate that the systematic discrepancy in
the vicinity of the primary minimum residuals (around $\varphi=-0.025$)  
is greatly reduced when
employing the new strategy. However, significant testing is required to
identify any potential problems with the new scheme. 

The second component of the residuals, the oscillatory pattern which can be spotted by eye in the original
curve, is certainly model-independent and is due either to the intrinsic 
variability or to an instrumental artifact. The latter hypothesis  is, however, unlikely, 
given the perfect phasing with the binary period over more than seven cycles. Furthermore,
none of the known frequencies of instrumental origin is close to  those found in the pattern  
components.
  \begin{figure}[!hb]
   \centering
     \includegraphics[width=8.7cm]{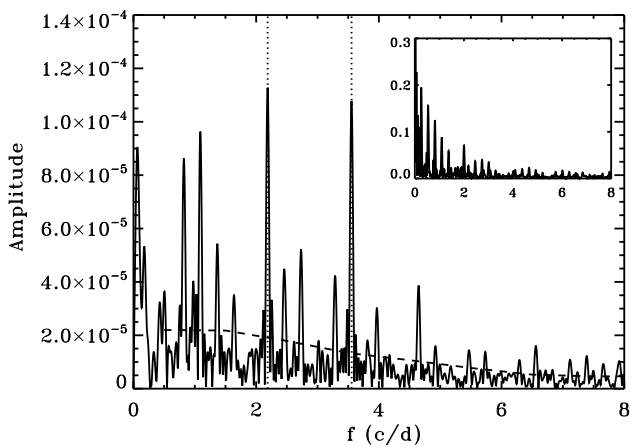}
     \caption{The Fourier spectrum of the  ``cleaned"  residuals after pre-withening  of the first
     two frequencies in Table \ref{freq_tab}. The dotted lines mark the two frequencies 
     8 $f_{\rm orb}$ and 13 $f_{\rm orb}$. The dashed line is the spectrum of noise, computed
     by Period04 on the residuals.
     The inset shows the spectral window.
 }
     \label{freqs}
	 \end{figure}
To extract the individual frequencies  we performed a Fourier analysis with Period04 \citep{p04}. 
For this purpose a complete set of residuals, in the time domain, was obtained by subtracting 
the fitted model from the $\sim$ 60000 points of the original light curve.
The analysis is made more complex  by the presence in the time series of residuals produced by 
the above-mentioned systematic effects, which generate power at the orbital frequency
($f_{\rm orb}=0.273444$  d$^{-1}$) and its overtones.
The analysis was therefore performed on two different time strings: the complete set of 
residuals and a ``cleaned" sub-sample, obtained by deleting the points the phase interval $-0.1\le \varphi \le 0.05$.  
Given the high total number of points it was possible as well to analyze only some
sections of the light curve residuals, e.g. each individual period.

All the analyses provide essentially the same result: a clear detection of two frequencies,
corresponding to 8 and 13 times the orbital frequency and marginal detections of other orbital
overtones. Table \ref{freq_tab} and Fig \ref{freqs} summarize the results of the analysis for
the ``cleaned" set of residuals. The first frequency corresponds to  $ f_{\rm orb}$ and was
kept fixed. 
The first two  frequencies are actually needed to fit the  slow trend, visible in the lower panel of 
Fig.~\ref{phased_lc}, due to the systematics introduced by  subtraction of  the light curve model.
 Similarly the low frequency F6  fits a long term residual trend,  
a left-over of the  detrending procedure which was applied  to the original curve. 
Frequencies F3 and F4 correspond to  $8 f_{\rm orb}$ and $ 13 f_{\rm orb}$.

The detections are very clear, reaching amplitudes over ten times the noise level. More multiples of the orbital frequency are present, but at lower amplitude so we consider them to be only marginal detections. 
Intrinsic variability at multiples of the orbital frequency are a
phenomenon pointing towards tidally induced oscillations or free
oscillations affected by the close binarity.

Even though several pulsating stars in close binaries have been found 
\citep[e.g.][]{aeha04}, oscillation frequencies at multiples of the orbital
frequency have only seldomly been detected.  The most likely explanation
is that they are due to resonant dynamic tides which can lead to detectable
amplitudes. As discussed by, e.g., \citet{wiae02}, the shape of
tidally induced observables (light curves or radial velocities) can vary from very
irregular for orbital periods away from a resonances with free
oscillation modes to sinusoidal for orbital periods close to a resonance with a free
oscillation mode.  It is somewhat surprising that we find clear multiples of orders
13 and 8 for HD\,174884, while the lower order multiples are not so markedly
found. Interestingly, such a case also occurs for the slowly pulsating B
star HD\,177863, where an oscillation frequency of exactly ten times the orbital
frequency was found and interpreted as due to resonant excitation \citep{decat00}.

According to theory, tides can efficiently excite the free oscillation modes of the star
which are close to the tidal frequency \citep[and its multiples in eccentric orbits see, e.g.,][]{wil03}.  
The dominant tidal term is associated with the spherical harmonics of degree  $\ell =2$.
Besides, given the typical values of the tidal frequencies in close systems, the stellar free oscillation
modes should be g-modes, and typically of high radial order, $n \gtrsim10$  \citep{zahn05}.

By means of the code LOSC \citep[Li\`{e}ge OScillation Code][]{Scuetal08b} 
we computed the eigenfrequencies of the primary star for all the
models within 2-$\sigma$ in mass and radius and $\Delta T_{\rm eff} \lesssim 1600$~K.
The purpose was to   check the properties of the $\ell=2$ modes close to
the F values of Table \ref{freq_tab}.
LOSC includes the effects of rotation on the frequencies in a perturbative approximation,
 and the correction terms are limited to the first order in $\omega_{\rm rot}$ (e.g. Ledoux 1951). 
 This approach is acceptable when the effects of the centrifugal force and of Coriolis acceleration are 
 small. The former  is estimated by the ratio  ($\epsilon$) between the rotational period and the dynamical time 
,  the latter by the ratio ($\mu$) of rotation to  oscillation frequencies. 
 For the \hd primary component  $\epsilon \sim 4\times 10^{-2}$ and $\mu=0.25$ for $F3$, and $\mu=0.15$ for $F4$.
  As shown by \citet{mmal08}, the effect of rotation on the internal chemical composition profiles can also
 lead to slight changes of the eigenfrequencies.
 
Since our aim is not to fit the observed frequencies, but to verify that their value is well in the 
range of $\ell=2$ g-modes with radial order $n$ of $\sim$10 or larger, we consider the
 first order approach  sufficient.
The comparison between observed and theoretical frequencies indicates, indeed, that  the lower frequency 
$F3$ corresponds,  for the models described in  Table \ref{models}, to  $n$ values between 8 and 10,
and the higher one to slightly larger values.
We can conclude, therefore, that the frequencies we find in the residuals are compatible with
tidally forced oscillations.

 As a further check of the light curve solution robustness, we pre-withened the original curve
by the two frequencies F3 and F4 and recomputed the best fit. The new solution is essentially
the same as the old one:  the largest difference between best-fit parameters before and after pre-withening
is   $0.4$\%.  The subtraction of the harmonic pattern does not substantially improve the 
correlation of residuals, because their dominant component, which we interpreted as due to model 
systematics, is essentially  left unchanged by the procedure.
\section{Summary and conclusions}
\label{conclusions}
This work has been carried out with two aims in mind:  to achieve a thorough
description of an  interesting binary and to test  the performance of  current 
binary star modeling on data  of unprecedented accuracy, as 
those available for the \corot{} seismo-field targets.
 \hd lends itself to both purposes:   its peculiar light curve suggests
 an unusual system configuration, worth a detailed study, and it is as well quite stable
 over the seven observed orbital periods so that,  if possible systematic errors 
 arising from  binary modeling were present, they would not be hidden by transient intrinsic phenomena. 
  (The latter is indeed the case of the other fully analyzed binary seismo-target of \corot\ first 
  runs, \object{AU Mon} 
 \citep{desmetal09} whose study reveals the presence of variable circumbinary material).
 
 Given the presence of a grazing eclipse, we were concerned about
  solution uniqueness, a problem which is sometimes overlooked, and we put a great effort in getting a
 sound estimate of parameter uncertainties, essential for sensible comparison with
 evolutionary stellar models.
   
 We can summarize our main results as follows:
 \begin{itemize}
 \item{due to the high  accuracy of CoRoT data, we were able to unambiguously derive parameters,
 such as inclination and component sizes, which would otherwise be poorly constrained in a system with
 grazing eclipses. This fact, the excellent agreement between the results of the different  methods applied 
for the analysis of the spectra, and the extensive check of solution uniqueness allowed a sound
estimation of the physical parameters of \hd  and their uncertainties.}
\item{The comparison with stellar  models  is quite satisfactory but for the temperature difference between
the components, which, for the best fitting models,  is typically a few hundred degrees higher than that 
from the light curve  solution.  We interpreted that as due to the comparison with non rotating stellar models.
We could as well improve the agreement with models introducing differences in the physical processes acting in the
components, such as overshooting, but Occam's razor arguments  favor  the simpler, though
not exhaustive, explanation. Increasing the free parameters yields, in fact, only a marginal improvement.
The comparison with theoretical models provided as well an estimation of the system age and
some indication on the component chemical composition.}  
 \item{The dynamical properties of \hd are  in very good agreement with the 
 predictions of Zahn's theory of circularization and spin-orbit synchronization. The high value
 of the eccentricity for the period suggests that  resonance locking has not been at work in
 the system.}    
 \item{A few frequencies are clearly detected in the residuals of the light curve fit at multiples
of the orbital frequency.  We tentatively interpret them as resonantly excited pulsations.
This hypothesis is strengthened by the fact that many g-modes of high radial order and degree 
$\ell=2$ exist in the  frequency range of  tidal frequencies.
  Among the very few detections of tidally induced pulsations in binaries,  \hd is the only case with 
two well characterized  early type MS components.}
 \end{itemize}

Finally, it is evident from  our results that when dealing with observation  accuracy of a few tenths  
of mmag  we are close to the level at which systematics from the model has a relevant impact on the 
light curve solution.
This indicates that the models shall be updated, especially in view of the still higher accuracy curves 
which will be obtained by current and future space missions such as  Kepler \citep{kepler}.
Similarly, the physical properties derived by data of such accuracy should be compared with 
theoretical models  including  stellar rotation. 


\begin{acknowledgements}
We thank  Arlette Noels, Franca D'Antona and Georges Meynet for enlighting discussions, 
Richard Scuflaire  and Andrea Miglio for providing the numerical codes, and John Southworth
for useful suggestions and comments on the manuscript.
 
 The research leading to these results has received funding from
 the Italian Space Agency (ASI) under  contract ASI/INAF I/015/07/00 in the frame of the ASI-ESS project,
from  the European Research Council under the European Community's Seventh 
Framework Programme (FP7/2007--2013)/ERC grant agreement n$^\circ$227224 (PROSPERITY),
as well as from the Research Council of K.U.Leuven grant agreement GOA/2008/04 and
from the Belgian PRODEX Office under contract C90309: CoRoT Data Exploitation.
 
EN was funded by means of a 6-month Visiting Postdoctoral Fellowship of the FWO,
Flanders in the framework of project G.0332.06 as well as by the European Helio-
and Asteroseismology Network (HELAS), a major international collaboration funded
by the European Commission's Sixth Framework Programme.
 The research of PH was supported by the grant 205/06/0304, and 205/08/H005
of the Czech Science Foundation and also from the Research Program
MSM0021620860 of the Ministry of Education of the Czech Republic.
AP acknowledges NSF/RUI Grant No. AST-05-07542.

\end{acknowledgements}

\bibliographystyle{aa} 
\bibliography{13311} 

\begin{thebibliography}{53}
\expandafter\ifx\csname natexlab\endcsname\relax\def\natexlab#1{#1}\fi

\bibitem[{{Aerts} \& {Harmanec}(2004)}]{aeha04}
{Aerts}, C. \& {Harmanec}, P. 2004, in Astronomical Society of the Pacific
  Conference Series, Vol. 318, Spectroscopically and Spatially Resolving the
  Components of the Close Binary Stars, ed. R.~W. {Hilditch}, H.~{Hensberge},
  \& K.~{Pavlovski}, 325--333

\bibitem[{{Asplund} {et~al.}(2005){Asplund}, {Grevesse}, \& {Sauval}}]{ags05}
{Asplund}, M., {Grevesse}, N., \& {Sauval}, A.~J. 2005, in Astronomical Society
  of the Pacific Conference Series, Vol. 336, Cosmic Abundances as Records of
  Stellar Evolution and Nucleosynthesis, ed. T.~G. {Barnes}, III \& F.~N.
  {Bash}, 25

\bibitem[{{Bastian} \& {de Mink}(2009)}]{bdm09}
{Bastian}, N. \& {de Mink}, S.~E. 2009, \mnras, 398, L11

\bibitem[{{Bevington} \& {Robinson}(2003)}]{bev03}
{Bevington}, P.~R. \& {Robinson}, D.~K. 2003, {Data reduction and error
  analysis for the physical sciences} ({McGraw-Hill Science Engineering})

\bibitem[{{B{\"o}hm-Vitense}(1958)}]{bv58}
{B{\"o}hm-Vitense}, E. 1958, Zeitschrift fur Astrophysik, 46, 108

\bibitem[{{Borucki} {et~al.}(2008){Borucki}, {Koch}, {Basri}, {Batalha},
  {Brown}, {Caldwell}, {Christensen-Dalsgaard}, {Cochran}, {Dunham}, {Gautier},
  {Geary}, {Gilliland}, {Jenkins}, {Kondo}, {Latham}, {Lissauer}, \&
  {Monet}}]{kepler}
{Borucki}, W., {Koch}, D., {Basri}, G., {et~al.} 2008, in IAU Symposium, Vol.
  249, IAU Symposium, ed. Y.-S. {Sun}, S.~{Ferraz-Mello}, \& J.-L. {Zhou},
  17--24

\bibitem[{{Castelli} \& {Kurucz}(2004)}]{castelli2004}
{Castelli}, F. \& {Kurucz}, R.~L. 2004, ArXiv Astrophysics e-prints

\bibitem[{{Claret} \& {Cunha}(1997)}]{clacu97}
{Claret}, A. \& {Cunha}, N.~C.~S. 1997, \aap, 318, 187

\bibitem[{{Claret} \& {Gimenez}(2005)}]{clagim05}
{Claret}, A. \& {Gimenez}, A. 2005, VizieR Online Data Catalog, 6118

\bibitem[{{Crawford} \& {Mandwewala}(1976)}]{crawman76}
{Crawford}, D.~L. \& {Mandwewala}, N. 1976, \pasp, 88, 917

\bibitem[{{De Cat} {et~al.}(2000){De Cat}, {Aerts}, {De Ridder}, {Kolenberg},
  {Meeus}, \& {Decin}}]{decat00}
{De Cat}, P., {Aerts}, C., {De Ridder}, J., {et~al.} 2000, \aap, 355, 1015

\bibitem[{{De Ridder} {et~al.}(2004){De Ridder}, {Telting}, {Balona},
  {Handler}, {Briquet}, {Daszy{\'n}ska-Daszkiewicz}, {Lefever}, {Korn},
  {Heiter}, \& {Aerts}}]{deridetal04}
{De Ridder}, J., {Telting}, J.~H., {Balona}, L.~A., {et~al.} 2004, \mnras, 351,
  324

\bibitem[{{Desmet} {et~al.}(2009){Desmet}, {Fremat}, {Baudin}, {Harmanec},
  {Lampens}, {Janot Pacheco}, {Briquet}, {Degroote}, {Neiner}, {Mathias},
  {Poretti}, {Rainer}, {Uytterhoeven}, {Amado}, {Valtier}, {Pr{\v s}a},
  {Maceroni}, \& {Aerts}}]{desmetal09}
{Desmet}, M., {Fremat}, Y., {Baudin}, F., {et~al.} 2009, \mnras, in press

\bibitem[{{Ferguson} {et~al.}(2005){Ferguson}, {Alexander}, {Allard}, {Barman},
  {Bodnarik}, {Hauschildt}, {Heffner-Wong}, \& {Tamanai}}]{fergal05}
{Ferguson}, J.~W., {Alexander}, D.~R., {Allard}, F., {et~al.} 2005, \apj, 623,
  585

\bibitem[{{Giuricin} {et~al.}(1984){Giuricin}, {Mardirossian}, \&
  {Mezzetti}}]{giuma84}
{Giuricin}, G., {Mardirossian}, F., \& {Mezzetti}, M. 1984, \aap, 134, 365

\bibitem[{{Gray} \& {Corbally}(1994)}]{grayco94}
{Gray}, R.~O. \& {Corbally}, C.~J. 1994, \aj, 107, 742

\bibitem[{{Grevesse} \& {Noels}(1993)}]{gn93}
{Grevesse}, N. \& {Noels}, A. 1993, in Origin and Evolution of the Elements,
  ed. N.~{Prantzos}, E.~{Vangioni-Flam}, \& M.~{Casse}, 15--25

\bibitem[{{Hadrava}(1995)}]{korel1}
{Hadrava}, P. 1995, \aaps, 114, 393

\bibitem[{{Hadrava}(1997)}]{korel2}
{Hadrava}, P. 1997, \aaps, 122, 581

\bibitem[{{Hadrava}(2004)}]{korel3}
{Hadrava}, P. 2004, Publ. Astron. Inst. Acad. Sci. Czech Rep., 92, 15

\bibitem[{{Heged{\"u}s} {et~al.}(2005){Heged{\"u}s}, {Gim{\'e}nez}, \&
  {Claret}}]{hege05}
{Heged{\"u}s}, T., {Gim{\'e}nez}, A., \& {Claret}, A. 2005, in Astronomical
  Society of the Pacific Conference Series, Vol. 333, Tidal Evolution and
  Oscillations in Binary Stars, ed. A.~{Claret}, A.~{Gim{\'e}nez}, \& J.-P.
  {Zahn}, 88

\bibitem[{{Hut}(1981)}]{hut1981}
{Hut}, P. 1981, \aap, 99, 126

\bibitem[{{Iglesias} \& {Rogers}(1996)}]{iglrog96}
{Iglesias}, C.~A. \& {Rogers}, F.~J. 1996, \apj, 464, 943

\bibitem[{{Kurucz}(1998)}]{kur98}
{Kurucz}, R.~L. 1998, Highlights of Astronomy, 11, 646

\bibitem[{{Lebreton} {et~al.}(2008){Lebreton}, {Montalb{\'a}n},
  {Christensen-Dalsgaard}, {Roxburgh}, \& {Weiss}}]{lebmon08}
{Lebreton}, Y., {Montalb{\'a}n}, J., {Christensen-Dalsgaard}, J., {Roxburgh},
  I.~W., \& {Weiss}, A. 2008, \apss, 316, 187

\bibitem[{{Lenz} \& {Breger}(2005)}]{p04}
{Lenz}, P. \& {Breger}, M. 2005, Communications in Asteroseismology, 146, 53

\bibitem[{{Maceroni} \& {Rucinski}(1997)}]{mr97}
{Maceroni}, C. \& {Rucinski}, S.~M. 1997, \pasp, 109, 782

\bibitem[{{Maeder} \& {Meynet}(2004)}]{maemey04}
{Maeder}, A. \& {Meynet}, G. 2004, in IAU Symposium, Vol. 215, Stellar
  Rotation, ed. {A.~Maeder \& P.~Eenens}, 500

\bibitem[{{Miglio} \& {Montalb{\'a}n}(2005)}]{mm05}
{Miglio}, A. \& {Montalb{\'a}n}, J. 2005, \aap, 441, 615

\bibitem[{{Miglio} {et~al.}(2007){Miglio}, {Montalb{\'a}n}, \&
  {Maceroni}}]{mmm07}
{Miglio}, A., {Montalb{\'a}n}, J., \& {Maceroni}, C. 2007, \mnras, 377, 373

\bibitem[{{Miglio} {et~al.}(2008){Miglio}, {Montalb{\'a}n}, {Noels}, \&
  {Eggenberger}}]{mmal08}
{Miglio}, A., {Montalb{\'a}n}, J., {Noels}, A., \& {Eggenberger}, P. 2008,
  \mnras, 386, 1487

\bibitem[{{Moon}(1985)}]{Moon85}
{Moon}, T.~T. 1985, Comm. University of London Observatory, 78

\bibitem[{{Moon} \& {Dworetsky}(1985)}]{moondwo85}
{Moon}, T.~T. \& {Dworetsky}, M.~M. 1985, \mnras, 217, 305

\bibitem[{{Napiwotzki} {et~al.}(1993){Napiwotzki}, {Schoenberner}, \&
  {Wenske}}]{napal93}
{Napiwotzki}, R., {Schoenberner}, D., \& {Wenske}, V. 1993, \aap, 268, 653

\bibitem[{{North} \& {Zahn}(2003)}]{northz03}
{North}, P. \& {Zahn}, J.-P. 2003, \aap, 405, 677

\bibitem[{{Press} {et~al.}(1992){Press}, {Teukolsky}, {Vetterling}, \&
  {Flannery}}]{NR}
{Press}, W.~H., {Teukolsky}, S.~A., {Vetterling}, W.~T., \& {Flannery}, B.~P.
  1992, {Numerical Recipes in FORTRAN, The Art of Scientific Computing}
  ({Cambridge University Press})

\bibitem[{{Pr{\v s}a} \& {Zwitter}(2005)}]{prsazw05}
{Pr{\v s}a}, A. \& {Zwitter}, T. 2005, \apj, 628, 426

\bibitem[{{Rieke} \& {Lebofsky}(1985)}]{riele85}
{Rieke}, G.~H. \& {Lebofsky}, M.~J. 1985, \apj, 288, 618

\bibitem[{{Rogers} \& {Nayfonov}(2002)}]{ronay02}
{Rogers}, F.~J. \& {Nayfonov}, A. 2002, \apj, 576, 1064

\bibitem[{{Rucinski} {et~al.}(2007){Rucinski}, {Kuschnig}, {Matthews},
  {Dimitrov}, {Pribulla}, {Guenther}, {Moffat}, {Sasselov}, {Walker}, \&
  {Weiss}}]{smr07}
{Rucinski}, S.~M., {Kuschnig}, R., {Matthews}, J.~M., {et~al.} 2007, \mnras,
  380, L63

\bibitem[{{Scuflaire} {et~al.}(2008{\natexlab{a}}){Scuflaire}, {Montalb{\'a}n},
  {Th{\'e}ado}, {Bourge}, {Miglio}, {Godart}, {Thoul}, \& {Noels}}]{Scuetal08b}
{Scuflaire}, R., {Montalb{\'a}n}, J., {Th{\'e}ado}, S., {et~al.}
  2008{\natexlab{a}}, \apss, 316, 149

\bibitem[{{Scuflaire} {et~al.}(2008{\natexlab{b}}){Scuflaire}, {Th{\'e}ado},
  {Montalb{\'a}n}, {Miglio}, {Bourge}, {Godart}, {Thoul}, \&
  {Noels}}]{scuetal08}
{Scuflaire}, R., {Th{\'e}ado}, S., {Montalb{\'a}n}, J., {et~al.}
  2008{\natexlab{b}}, \apss, 316, 83

\bibitem[{{Skrutskie} {et~al.}(2006){Skrutskie}, {Cutri}, {Stiening},
  {Weinberg}, {Schneider}, {Carpenter}, {Beichman}, {Capps}, {Chester},
  {Elias}, {Huchra}, {Liebert}, {Lonsdale}, {Monet}, {Price}, {Seitzer},
  {Jarrett}, {Kirkpatrick}, {Gizis}, {Howard}, {Evans}, {Fowler}, {Fullmer},
  {Hurt}, {Light}, {Kopan}, {Marsh}, {McCallon}, {Tam}, {Van Dyk}, \&
  {Wheelock}}]{2mass}
{Skrutskie}, M.~F., {Cutri}, R.~M., {Stiening}, R., {et~al.} 2006, \aj, 131,
  1163

\bibitem[{{Solano} {et~al.}(2005){Solano}, {Catala}, {Garrido}, {Poretti},
  {Janot-Pacheco}, {Guti{\'e}rrez}, {Gonz{\'a}lez}, {Mantegazza}, {Neiner},
  {Fremat}, {Charpinet}, {Weiss}, {Amado}, {Rainer}, {Tsymbal}, {Lyashko},
  {Ballereau}, {Bouret}, {Hua}, {Katz}, {Ligni{\`e}res}, {L{\"u}ftinger},
  {Mittermayer}, {Nesvacil}, {Soubiran}, {van't Veer-Menneret}, {Goupil},
  {Costa}, {Rolland}, {Antonello}, {Bossi}, {Buzzoni}, {Rodrigo}, {Aerts},
  {Butler}, {Guenther}, \& {Hatzes}}]{sol05}
{Solano}, E., {Catala}, C., {Garrido}, R., {et~al.} 2005, \aj, 129, 547

\bibitem[{{Southworth} \& {Clausen}(2007)}]{southworth2007}
{Southworth}, J. \& {Clausen}, J. 2007, \aap, 461, 1077

\bibitem[{{Stellingwerf}(1978)}]{stell78}
{Stellingwerf}, R.~F. 1978, \apj, 224, 953

\bibitem[{{van Leeuwen}(2007)}]{vanle07}
{van Leeuwen}, F. 2007, \aap, 474, 653

\bibitem[{{von Zeipel}(1924)}]{vonzeipel1924}
{von Zeipel}, H. 1924, \mnras, 84, 665

\bibitem[{{Waelkens} {et~al.}(1998){Waelkens}, {Aerts}, {Kestens}, {Grenon}, \&
  {Eyer}}]{wae98}
{Waelkens}, C., {Aerts}, C., {Kestens}, E., {Grenon}, M., \& {Eyer}, L. 1998,
  \aap, 330, 215

\bibitem[{{Willems}(2003)}]{wil03}
{Willems}, B. 2003, \mnras, 346, 968

\bibitem[{{Willems} \& {Aerts}(2002)}]{wiae02}
{Willems}, B. \& {Aerts}, C. 2002, \aap, 384, 441

\bibitem[{{Witte} \& {Savonije}(1999)}]{wittesav99}
{Witte}, M.~G. \& {Savonije}, G.~J. 1999, \aap, 350, 129

\bibitem[{{Zahn}(2005)}]{zahn05}
{Zahn}, J.-P. 2005, in Astronomical Society of the Pacific Conference Series,
  Vol. 333, Tidal Evolution and Oscillations in Binary Stars, ed. A.~{Claret},
  A.~{Gim{\'e}nez}, \& J.-P. {Zahn}, 4--23

\end{thebibliography}
\end{document}